\documentclass[pre,aps,showpacs,groupedaddress]{revtex4-2}
\usepackage{graphicx}
\usepackage{amssymb}
\usepackage{textcomp}
\usepackage{amsmath}

\begin{document}

%\title{Non-Gaussian aspects of deep and wide trigonometric networks}
\title{Bayesian inference with finitely wide neural networks}

\author{Chi-Ken~Lu}
\email{Faculty of Data Science Minor, CL1178@rutgers.edu}
\affiliation{Department of Mathematics and Computer Science, Rutgers University, Newark, New Jersey 07102, USA}

\date{\today}

\begin{abstract}

The analytic inference, e.g. predictive distribution being in closed form, may be an appealing benefit for machine learning practitioners when they treat wide neural networks as Gaussian process in Bayesian setting. The realistic widths, however, are finite and cause weak deviation from the Gaussianity under which partial marginalization of random variables in a model is straightforward. On the basis of multivariate Edgeworth expansion, we propose a non-Gaussian distribution in differential form to model a finite set of outputs from a random neural network, and derive the corresponding marginal and conditional properties. Thus, we are able to derive the non-Gaussian posterior distribution in Bayesian regression task. In addition, in the bottlenecked deep neural networks, a weight space representation of deep Gaussian process, the non-Gaussianity is investigated through the marginal kernel and the accompanying small parameters.

\end{abstract}

%\pacs{73.43.Cd,03.67.Mn,73.43.Nq}

\maketitle

\section{Introduction}

Neal in his seminal work~\cite{neal1997monte} pointed out that a shallow but infinitely wide random neural network is a Gaussian process (GP)~\cite{rasmussen2006gaussian} in statistical sense. 
%This also suggests a profound connection between the gradient-based learning of random neural network in weight space and the Bayesian learning in function space~\cite{jacot2018neural,arora2019exact}. 
Subsequent work~\cite{williams1997computing,cho2009kernel} in interpreting neural network with specific nonlinear activation units as kernel machines was also inspired by such idea. More recent reports~\cite{matthews2018gaussian,lee2018deep} further claimed the equivalence between GP and deep neural networks when each hidden layer in latter is of infinite width. Consequently, machine learning practitioners can perform Bayesian inference by treating deep and wide neural network as a GP, and exploit the analytic marginal and conditional properties of multivariate Gaussian distribution. Otherwise, one needs to employ gradient-based learning and bootstrap sampling for obtaining predictive distribution~\cite{osband2016risk}.

%The role of large widths is suppressing the fluctuation of covariance between network outputs, resulting in fixed kernel functions which only depend on activation units and hyperparameters. More recently, Ref. concluded that the argument is still valid for the models of deep neural networks as long as each of the hidden layer is infinitely wide. For  using deep and wide neural networks as model, the issue of robustness of inference and uncertainty estimation can be resolved once the networks are treated as GP. Alternatively,  

In reality, all neural networks have finite width. Therefore, the deviation from Gaussianity requires further quantitative account as practitioners may wonder the corrections to the predictive mean and variance in, for example, a regression task.
%which poses challenges for understanding random deep neural networks in function space without the help of fixed kernel. On the other hand, without converging to the fix kernel, the networks with finite width are more capable of learning representation of data than their infinitely wide counterparts~\cite{lee2020finite,aitchison2020bigger}. 
Yaida~\cite{yaida2020non} and colleagues~\cite{roberts2021principles} proposed a perturbative approach for computing the multivariate cumulants by direct application of Wick's contraction theorem. Moreover, the fourth cumulants are shown to be nonzero, scaled by the sum of reciprocal widths, $1/N_1+1/N_2+\cdots+1/N_{L-1}$, and signaling non-Gaussian aspect of the random processes representing finite-width deep neural networks with $L$ hidden layers~\cite{yaida2020non}. A quartic energy functional for a fixed set of network outputs is formulated field-theoretically with which the corrections to the posterior mean and variance due to the {\it weak} non-Gaussianity are obtained~\cite{yaida2020non}.  

While Yaida's approach is appealing from a field theory perspective (see also~\cite{dyer2019asymptotics,gabrie2020mean}), the loss of elegant marginal property due to the presence of fourth power term in exponent is critical for analytic inference with finite-width networks. An alternative thinking is to modify the multivariate Gaussian distribution so that the new distribution can match the higher cumulants associated with the networks. In this paper, we shall use the   
%The computation of cumulants is tedious but straightforward with application of Wick's contraction theorem. 
%We propose an alternative approach which recursively approximate the hidden layer's output distribution using the 
multivariate Edgeworth series expansion~\cite{skovgaard1986multivariate} to construct the non-Gaussian distribution for the network's outputs. In particular, we find that the differential representation of Edgeworth expansion greatly facilitates the derivation of marginal and conditional properties of the non-Gaussian distribution. Three main results are reported in this paper. First, the marginal property is intact, and the corrections to conditional mean and variance are derived. Second, with observed data, the non-Gaussian posterior distribution associated with an unobserved output is derived. Third, we derive the marginal covariance~\cite{lu2020interpretable} of a bottlenecked deep neural network~\cite{agrawal2020wide}, which represents deep Gaussian Process~\cite{damianou2013deep} in weight space. It is worthwhile to note that some of the hidden layers in the bottlenecked network are narrow and {\it strong} non-Gaussianity may be induced.

The paper has the following organization. In Sec.~\ref{network} we begin by reviewing the computational structure of a shallow cosine network, its equivalence to GP with a Gaussian kernel, and the emergence of non-Gaussianity due to finite width. The shallow network with random parameters in Bayesian setting is a non-Gaussian prior over function. In Sec.~\ref{edgeworth} the non-Gaussian prior is represented as a differential representation of Edgeworth expansion around Gaussian multivariate distribution, and its marginal and conditional properties are established. Application of the non-Gaussian prior in Bayesian regression task is discussed in Sec.~\ref{Bayesian}. Finally, Sec.~\ref{bottleneck} is devoted to investigating the combined effect of nonlinear activation, depth and finite width on the non-Gaussian prior for a deep bottlenecked cosine network, which is followed by a discussion in Sec.~\ref{discussion}.
%For cumulants up to the fourth order, the fluctuating kernels due to the finite width are taken into account. More importantly, the {\it strong} non-Gaussianity arising in the family of bottlenecked deep neural networks~\cite{agrawal2020wide} due to the narrow hidden layers can be captured. In addition, the bottlenecked networks can be considered as weight space representation of , and the depth can cause pathological issues for learning~\cite{duvenaud2014avoiding,dunlop2018deep}, which is also discussed in the context of deep neural network~\cite{roberts2021principles}.     

\section{Wide feed forward neural network}\label{network}

Let us start with the discussion of a random shallow feed forward network with cosine activation. In the infinite width limit, the network is statistically equivalent to a GP with Gaussian kernel~\cite{rahimi2008random}. The goal is to see the emergence of nonzero fourth cumulant when the width is finite. Consider a single output network with $N$ activation units, the real function value $z_\alpha\in\mathbb R$ with greek subscript is an indexed random variable in association with its input ${\bf x}_{\alpha}\in\mathbb R^d$. Explicitly, the output-input of the network has the following relation,
\begin{equation}
	z_{\alpha}=\sqrt{\frac{2}{N}}\sum_{i=1}^{N}w_{i}\cos(
	\frac{\Omega_i\cdot{\bf x}_{\alpha}}{\sqrt{d}}
	+\phi_i)\:,\label{shallow_network}
\end{equation} where the weight variables $w$'s are sampled from the Gaussian distribution $\mathcal N(0,1)$, the scaling variables $\Omega$'s sampled from $\mathcal N(0,I_d)$, and the phase variables $\phi$'s sampled from uniform distribution $\mathcal U([0,2\pi])$. The normalization factors $\sqrt{2/N}$ and $1/\sqrt{d}$ above follow the parameterization used in~\cite{jacot2018neural}. Because $w$ is zero-mean, the first relevant statistical moment is the covariance, $k({\bf x}_{\alpha},{\bf x}_{\beta})$, namely the expectation of product of two function values at two inputs, 
\begin{equation}
\begin{split}
k({\bf x}_{\alpha},{\bf x}_{\beta})&:=\mathbb E[z_{\alpha}z_{\beta}]\\%\mathbb E\big\{
	%\exp[-\frac{1}{H}\sum_{j=1}^H
	%(z_{j,\alpha}^{(l-1)}-z_{j,\beta}^{(l-1)})^2]
	&=\frac{1}{N}\sum_{i}\mathbb E\bigg\{
	\cos[\frac{\Omega_i\cdot({\bf x}_{\alpha}-{\bf x}_{\beta})}
	{\sqrt{d}}]+
	\cos[\frac{\Omega_i\cdot({\bf x}_{\alpha}+{\bf x}_{\beta})}
	{\sqrt{d}}+2\phi_i]
	\bigg\}\\
	&=e^{-\frac{1}{2d}|{\bf x}_{\alpha}-{\bf x}_{\beta}|^2}\:.\label{2nd_moment}
	%\big\}
\end{split}
\end{equation} In above, the independence between the random variables $w$'s is used to arrive at the second equality. The third equality is due to the fact that Fourier transformation of Gaussian is Gaussian and the vanishing average of cosine with uniformly random phase. The fourth moment can be computed in similar manners, but one shall notice that the product of two cosine terms containing the random phase can generate a nonzero contribution. Here, we focus on the fourth cumulant tensor as the signature of non-Gaussianity,
\begin{equation}
\begin{split}
	V_{\alpha\beta\gamma\delta}&:=\mathbb E[z_{\alpha}z_{\beta}z_{\gamma}z_{\delta}]
	-K_{\alpha\beta}K_{\gamma\delta}-K_{\alpha\gamma}K_{\beta\delta}-K_{\alpha\delta}K_{\beta\gamma}\\
	&=\frac{1}{N^2}\sum_{ij}\mathbb E\big\{
	\cos[\frac{\Omega_i\cdot({\bf x}_{\alpha}+{\bf x}_{\beta})}
	{\sqrt{d}}+2\phi_i]
	\cos[\frac{\Omega_j\cdot({\bf x}_{\gamma}+{\bf x}_{\delta})}
	{\sqrt{d}}+2\phi_j]
	\big\} + {\rm sym.\ perm.}\\
	&=\frac{1}{2N}\bigg(
	%\mathbb E\big \{
		%e^{-\frac{1}{2}|{\bf z}^{(l-1)}_{\alpha}-{\bf z}^{(l-1)}_{\beta}|^2}
		%e^{-\frac{1}{2}|{\bf z}^{(l-1)}_{\gamma}-{\bf z}^{(l-1)}_{\delta}|^2}
		e^{-\frac{1}{2d}|{\bf x}_{\alpha}+{\bf x}_{\beta}
		-{\bf x}_{\gamma}-{\bf x}_{\delta}|^2}+
		e^{-\frac{1}{2d}|{\bf x}_{\alpha}+{\bf x}_{\gamma}
		-{\bf x}_{\beta}-{\bf x}_{\delta}|^2}+
		e^{-\frac{1}{2d}|{\bf x}_{\alpha}+{\bf x}_{\delta}
		-{\bf x}_{\beta}-{\bf x}_{\gamma}|^2}
		%\big \}
	\bigg)\:.\label{4th_cumulant}
\end{split}
\end{equation} Hence, the fourth cumulant tensor receives a contribution proportional to the reciprocal width, $1/N$, and it is symmetric with respect to permutation of indices.

\section{Multivariate Edgeworth expansion and non-Gaussian prior}\label{edgeworth}

The fourth cumulant tensor in Eq.~(\ref{4th_cumulant}) signifies that the distribution over the network outputs is non-Gaussian. In other words, the random parameters $w$'s, $\Omega$'s, and $\phi$'s from a prior distribution induce the non-Gaussian prior function distribution due to the finite width. The first interesting consequence of the nonzero fourth cumulant is that the prior distribution over a single output, say $z_{\alpha}$, receives a correction term deviating from Gaussian~\cite{antognini2019finite},
\begin{equation}
	q(z_{\alpha})=\mathcal N(z_{\alpha}|0,\sigma_0^2)\bigg[
	1+\frac{V_{\alpha\alpha\alpha\alpha}}{24}(
	3-6\frac{z_{\alpha}^2}{\sigma_0^2}+
	\frac{z_{\alpha}^4}{\sigma_0^4})
	\bigg]\:.
\end{equation} The variance $\sigma_0^2:=k(z_{\alpha},z_{\alpha})$ and the new distribution $q$ has matched moments, i.e. $\mathbb E_q[1]=1$, $\mathbb E_q[z_{\alpha}^2]=\sigma_0^2$, and $\mathbb E_q[z_{\alpha}^4]=3\sigma_0^4+V_{\alpha\alpha\alpha\alpha}$. Although the work~\cite{antognini2019finite} derived the above non-Gaussian distribution from a renormalization group perspective, such expansion with Hermite polynomials around Gaussian distribution has been known as Edgeworth expansion in statistics literature~\cite{papoulis2002probability,welling2005robust}. When the width $N\rightarrow\infty$, the fourth cumulant $V$ vanishes and for notational convenience we denote the limiting distribution by $q_{\infty}$, which is Gaussian. 
%The polynomial correction terms within the parenthesis 
%In the followings, we shall 

\subsection{Joint distribution}

As we are mainly interested in the prior distribution over a finite set of function values $\{z_1,z_2,\cdots\}$, the objective here is to construct the non-Gaussian joint distribution as a prior in Bayesian learning. The work~\cite{skovgaard1986multivariate} suggested the construction of multivariate Edgeworth expansion by replacing the univariate Gaussian $\mathcal N(z|0,\sigma_0^2)$ with multivariate one $\mathcal N({\bf z}|0,K)$ and the Hermite polynomials with contracted tensor terms containing the inverse covariance matrix $[K^{-1}]$ and vector of function values ${\bf z}$. Please see Appendix~\ref{Edgeworth_Expansion} for the expression.

An apparent difficulty with such representation is the demonstration of consistency after marginalization, i.e., $\int dz_1q(z_1,z_2,z_3,\cdots)=q(z_2,z_3,\cdots)$, which is of critical importance for deriving conditional distribution and subsequent Bayesian learning. Inspired by the fact that the Hermite polynomials are obtained by taking derivative of Gaussian, we can rewrite the multivariate Edgeworth expansion in the following differential form, 
\begin{equation}
	q({\bf z})=\big(
	1+\frac{V_{ijml}}{24}\partial_{z_i}\partial_{z_j}\partial_{z_m}\partial_{z_l}
	\big)
	\mathcal N({\bf z}|0,K)\label{Edgeworth_Joint}
\end{equation} where the notation of summation by repeated indices is employed. An advantage of such representation is that the adjoint property of derivative, $\partial_z^{\dag}=-\partial_z$, can be exploited in the integration of functions which vanish at infinity. %treated as Due to the form of derivative, 
Hence, it is easy to see the distribution $q$ is normalized, and the second and fourth moments match, for instance,
\begin{equation}
\begin{split}
	\mathbb E_q[z_1z_2]&=K_{12}\:,\\
%\end{equation} for arbitrary $a,b\in[N]$, and
%\begin{equation}
	\mathbb E_q[z_1z_2z_3z_4] &= K_{12}K_{34} + K_{13}K_{24} + K_{14}K_{23} + V_{1234}\:.
\end{split}
\end{equation}

The above non-Gaussian prior can be viewed as a perturbative extension of multivariate Gaussian. In the setting of Bayesian regression, the marginal and conditional properties of Gaussian are appealing when conjugate likelihood functions are employed. In the following, we shall examine the effects of the perturbation in differential form on these properties.     

\subsection{Marginal consistency and conditional statistics}
We first show that the marginal consistency is intact with the Edgeworth expansion in differential form. Namely, without loss of generality, it suffices to show that 
%The approximate distribution $q$ satisfies the marginal condition, namely, following its definition, 
marginalization of $z_1$ of the joint distribution $q(z_1,z_2,z_3,\cdots)$, %which can be proceeded as followings,%leads to the the following relations,
\begin{equation}
\begin{split}
	\int dz_1q(z_1,\tilde{\bf z}) &= \mathcal N(\tilde{\bf z}|0,\tilde K)+
	v_{\tilde{i}\tilde{j}\tilde{m}\tilde{l}}\partial_{z_{\tilde i}}\partial_{z_{\tilde j}}\partial_{z_{\tilde m}}\partial_{z_{\tilde l}}
	\int dz_1\mathcal N(z_1,\tilde{\bf z}|0,K)\\
	&= \big(
	1+v_{\tilde{i}\tilde{j}\tilde{m}\tilde{l}}\partial_{z_{\tilde i}}\partial_{z_{\tilde j}}\partial_{z_{\tilde m}}\partial_{z_{\tilde l}}
	\big)\mathcal N(\tilde{\bf z}|0,\tilde K)\\
	&=q(\tilde{\bf z})\:,
\end{split}
\end{equation} indeed reproduces the joint distribution for $z_2$, $z_3,\cdots$ in the original form Eq.~(\ref{Edgeworth_Joint}) as
the smaller covariance matrix $\tilde K$ is the submatrix of $K$ excluding the first row and column. For simplicity, we use $v$ to denote $V/24$. In deriving above, we have employed the fact that $\int dz_1(\partial_{z_1})^s\mathcal N(z_1,\tilde{\bf z})=0$ for any integer power $s\geq 1$, meaning that the indices $i,j,m,l$ shall exclude any contribution associated with $z_1$. Thus, the tilded indices $\tilde i\in\{2,3,4,\cdots\}$, and they can be factored out of the integral.

Here we wish to stress that the above marginal property of non-Gaussian prior may seems straightforward with the differential representation of Edgeworth expansion, but it does not seem promising to reach the same conclusion if the marginalization is carried out with the explicit representation in Appendix~\ref{Edgeworth_Expansion}. Nevertheless, one can easily write the conditional distribution $q(z_1|\tilde{\bf z})=q(z_1,\tilde{\bf z})/q(\tilde{\bf z})$. More interestingly,
%We can continue taking advantage of the differential representation. 
the conditional mean is useful for prediction with noiseless observation of data, and also shed light on the effect of non-Gaussianity due to the finite width. The details of deriving the following conditional mean %can be found in Appendix
%for the conditional distribution $q(z_1|\tilde{\bf z})$,
\begin{equation}
\begin{split}
	\mathbb E[z_1|\tilde{\bf z}]&=\int dz_1z_1\frac{q(z_1,\tilde{\bf z})}{q(\tilde{\bf z})}\\
	%&=\frac{1}{q(\tilde{\bf z})}\bigg[
	%\mu(\tilde {\bf z}) +
	%\frac{V_{\tilde{i}\tilde{j}\tilde{m}\tilde{l}}}{24}
	%\partial_{z_{\tilde i}}\partial_{z_{\tilde j}}\partial_{z_{\tilde m}}\partial_{z_{\tilde l}}
	%\mu(\tilde {\bf z}) -
	%%\mu(1+v_{\tilde{i}\tilde{j}\tilde{m}\tilde{l}}\partial_{z_{\tilde i}}\partial_{z_{\tilde j}}\partial_{z_{\tilde m}}\partial_{z_{\tilde l}})
	%\frac{V_{1\tilde{j}\tilde{m}\tilde{l}}}{6}
	%\partial_{z_{\tilde j}}\partial_{z_{\tilde m}}\partial_{z_{\tilde l}}
	%\bigg]\mathcal N(\tilde{\bf z}|0,\tilde K)\\
	&=\mu(\tilde{\bf z})+\frac{1}{q(\tilde{\bf z})}
	A_{\tilde{i}\tilde{j}\tilde{m}}
	%\bigg(
	%\frac{V_{\tilde{i}\tilde{j}\tilde{m}\tilde{l}}}{6}[{\tilde K}^{-1}{\bf k}]_{\tilde{l}}
	%-\frac{V_{1\tilde{i}\tilde{j}\tilde{m}}}{6}
	%\bigg)
	\partial_{z_{\tilde i}}\partial_{z_{\tilde j}}\partial_{z_{\tilde m}}
	\mathcal N(\tilde{\bf z}|0,\tilde K)\:,\label{pred_mean}
\end{split}
\end{equation} with the finite-width correction term proportional to the third-order tensor,  
\begin{equation}
	 A_{\tilde{i}\tilde{j}\tilde{m}}=
	%\bigg(
	\frac{V_{\tilde{i}\tilde{j}\tilde{m}\tilde{l}}}{6}[{\tilde K}^{-1}{\bf k}]_{\tilde{l}}
	-\frac{V_{1\tilde{i}\tilde{j}\tilde{m}}}{6}\:,
	%\bigg)
\end{equation} can be found in Appendix~\ref{cond-mean}.
%in which the appearance of $z_1$ in the integration allows no more than one of the indices $i,j,m,l$ to include $z_1$, and the adjoint of $\partial_{z_1}$ is used to arrive at the second equality. 
In the GP limit, the conditional mean becomes the well-known result $\mu(\tilde{\bf z})={\bf k}^t{\tilde K}^{-1}{\tilde{\bf z}}$, as the fourth cumulant $V$ vanishes in large $N$ limit. We also remind the readers that the tilded symbols are associated with the conditioned outputs $z_{2,3,\cdots}$.
%for Gaussian appears as the contribution when none of the indices involves $z_1$. 
In a similar manner, we can also show the conditional second moment,
\begin{equation}
\begin{split}
	\mathbb E[z_1^2|\tilde{\bf z}]&=\int dz_1z_1^2\frac{q(z_1,\tilde{\bf z})}{q(\tilde{\bf z})}\\
	%&=\sigma^2+
	%\frac{1}{q(\tilde{\bf z})}\bigg[\mu^2(\tilde{\bf z})+
	 %\frac{V_{\tilde{i}\tilde{j}\tilde{m}\tilde{l}}}{24}
	 %\partial_{z_{\tilde i}}\partial_{z_{\tilde j}}\partial_{z_{\tilde m}}\partial_{z_{\tilde l}}
	 %\mu^2(\tilde{\bf z})
	%%(1+v_{\tilde{i}\tilde{j}\tilde{m}\tilde{l}}\partial_{z_{\tilde i}}\partial_{z_{\tilde j}}\partial_{z_{\tilde m}}\partial_{z_{\tilde l}})
	%- \frac{V_{1\tilde{j}\tilde{m}\tilde{l}}}{3}\partial_{z_{\tilde j}}\partial_{z_{\tilde m}}\partial_{z_{\tilde l}}\mu(\tilde{\bf z})
	%+ \frac{V_{11\tilde{m}\tilde{l}}}{2}
	%\partial_{z_{\tilde m}}\partial_{z_{\tilde l}}
	%\bigg]\mathcal N(\tilde{\bf z}|0,\tilde K)\\
	&=\mu^2(\tilde{\bf z})+\sigma^2+\frac{1}{q(\tilde{\bf z})}\bigg[
	2\mu(\tilde{\bf z})A_{\tilde{i}\tilde{j}\tilde{m}}
	%(\frac{V_{\tilde{i}\tilde{j}\tilde{m}\tilde{l}}}{3}[{\tilde K}^{-1}{\bf k}]_{\tilde{l}}
	%-\frac{V_{1\tilde{i}\tilde{j}\tilde{m}}}{3})
	\partial_{z_{\tilde i}}\partial_{z_{\tilde j}}\partial_{z_{\tilde m}}+
	B_{\tilde{i}\tilde{j}}\partial_{z_{\tilde i}}\partial_{z_{\tilde j}}%\\
	%&(\frac{V_{\tilde{i}\tilde{j}\tilde{m}\tilde{l}}}{3}[{\tilde K}^{-1}{\bf k}]_{\tilde{l}})
	\bigg]\mathcal N(\tilde{\bf z}|0,\tilde K)
	\:,\label{pred_2nd_moment}
\end{split}
\end{equation} with the second-order tensors denoting
\begin{equation}
	B_{\tilde{i}\tilde{j}} = \frac{V_{\tilde{i}\tilde{j}\tilde{m}\tilde{l}}}{2}[{\tilde K}^{-1}{\bf k}]_{\tilde{m}}
	[{\tilde K}^{-1}{\bf k}]_{\tilde{l}}
	-V_{1\tilde{i}\tilde{j}\tilde{m}}[{\tilde K}^{-1}{\bf k}]_{\tilde{m}} +
	\frac{V_{11\tilde{i}\tilde{j}}}{2}\:.
\end{equation}
The conditional second moment coincides with that in the GP limit, $\sigma^2=K_{11}-{\bf k}^t{\tilde K}^{-1}{\bf k}$. The details of derivation can also be found in Appendix~\ref{cond-mean}. 
%The above can be understood as: in the second line, the first term appears due to the exclusion of contribution related to any power of $\partial_{z_1}$. The first and second terms within the square bracket arise as $\partial_{z_1}$ and $\partial^2_{z_1}$, respectively, are involved. 
It is worthwhile to note that the conditional variance, $\mathbb E_q[z_1^2|\tilde{\bf z}]-(\mathbb E_q[z_1|\tilde{\bf z}])^2$, can include contributions related to function values $\tilde{\bf z}$ through the nonzero $A$ and $B$ terms in above expressions. The lack of such dependence in GP limit is in fact a shortcoming for modelling, which has motivated the study of non-Gaussian prior in machine learning community (for example the Student-t process in~\cite{shah2014student}). 

\subsection{An example: bivariate distribution and prediction}

Now let us pause to consider a simple example where a shallow network defined in Eq.~(\ref{shallow_network}) with width $N$ is used to model two input-output pairs $({\bf x}_1, z_1)$ and $({\bf x}_2, z_2)$. In the end, we shall present the predictive mean and variance for $z_1$ conditioning on $z_2$. The following notations are used for simplification: the covariance 
%between the inputs is 
$c:=\exp(-|{\bf x}_1-{\bf x}_2|^2/2d)$, %due to the cosine activation and Gaussian network parameters. Before we explore their non-Gaussian bivariate distribution and subsequent inference, let us first introduce the notations related to the derivative of bivariate Gaussian with 
the 2-by-2 covariance matrix $\Sigma=\bigl( \begin{smallmatrix}1 & c\\ c & 1\end{smallmatrix}\bigr)$ for the bivariate prior, and $z_{21}:=(z_2-cz_1)/(\sqrt{1-c^2})$. Besides, the relations of derivatives of Gaussian in terms of Hermite polynomial shall be useful, 
%If $z_2$ is known, what is the distribution $p(y_1|y_2)$ over the unobserved $y_1$? Following above discussion and exploit the fact 
\begin{equation}
	\partial^n_{z_1}\partial^m_{z_2}\mathcal N(z_1,z_2|0,\Sigma)\propto\partial^n_{z_1}
	\big[
	\mathcal N(z_1|0,1)\partial^m_{z_2}\mathcal N(z_{21}|0,1)
	\big]\propto(-1)^m
	\frac{\mathcal N(z_2|0,1)}{\sqrt{(1-c^2)^m}}
	\partial^n_{z_1}[H_m(z_{21})\mathcal N(z_{12}|0,1)]\:,
\end{equation} where we have used %$z_{21}:=(z_2-cz_1)/(\sqrt{1-c^2})$ to 
$\mathcal N(z_1,z_2|0,\Sigma)\propto\mathcal N(z_1|0,1)\mathcal N(z_{21}|0,1)$ up to some irrelevant constant. The probabilist's Hermite polynomials are defined as $H_n(z)=(-1)^n[\mathcal N(z|0,1)]^{-1}d^n_z\mathcal N(z|0,1)$~\cite{papoulis2002probability}. Consequently, following the previous discussion, the non-Gaussian bivariate distribution is shown to be%Similar notations result in %and $z_{12}=(z_1-cz_2)/(\sqrt{1-c^2})$
\begin{equation}
\begin{split}
	q(z_1,z_2)&=\mathcal N(z_1,z_2|0,\Sigma)\bigg\{
	1+\frac{\gamma^4}{16N}[H_4(z_{21})+H_4(z_{12})]
	+\frac{c\gamma^4}{4N}[3c H_2(z_{21})+H_3(z_{21})H_1(z_{12})+z_{12}\leftrightarrow z_{21}]\\
	&+\frac{(c^4+2)\gamma^4}{8N}[2c^2 + 4c H_1(z_{21})H_1(z_{12})+H_2(z_{21})H_2(z_{12})]
	\bigg\}\:,
\end{split}
\end{equation} and the symbol $\gamma=1/\sqrt{1-c^2}$ is used to simplify the expression. Besides, the fourth cumulant for the cosine network, $V_{1111}=3/2N$, $V_{1112}=3c/2N$, and $V_{1122}=(c^4+2)/2N$, have been plugged in above. As for the conditional distribution, e.g. $q(z_1|z_2)=q(z_1,z_2)/q(z_2)$, for the bivariate case, one can simply divid the above with $q(z_2)=\mathcal N(z_2|0,1)[1+H_4(z_2)/16N]$, which shall yield a conditional Gaussian $\mathcal N(z_1|z_2)$ multiplied by a factor representing the finite-width effect. 

%\begin{equation}
%\begin{split}
%	q(z_1|z_2)&=\frac{q(z_1,z_2)}{q(z_2)}\\
%	&=\frac{1}
%	{1+\frac{V_{2222}}{24}He(z_2)}
%\end{split}
%\end{equation}
%Besides the Monte Carlo approach (e.g. see Ch.~44 in~\cite{mackay2003information}) from the weight space perspective, 
%One can just take the infinite width limit and then apply the expressions of GP inference, obtaining the distribution with noiseless assumption, 
%\begin{equation*}
%	p(y_1|y_2)=\mathcal N(y_1|\mu_{1|2},\sigma_{1|2}^2)\:,
%\end{equation*} where $\mu_{1|2}=cy_2$ and $\sigma_{1|2}^2=1-c^2$. Here, $c$ denotes the expectation of covariance $k({\bf x}_1, {\bf x}_2)=\exp(|{\bf x}_1-{\bf x}_2|^2/2d)$ from Eq.(\ref{2nd_moment}) between the inputs. Due to the Gaussianity, the predictive mean for $y_1$ depends linearly on $y_2$ and only the inputs enter the predictive variance. Now we proceed to ask what are the effects of non-Gaussian prior induced by the finite width? The distribution becomes $p(y_1|y_2)=q(y_1,y_2)/q(y_2)$, and, with some effort,
%\begin{equation}
%	p(y_1|y_2)=\mathcal N(y_1|\mu_{1|2},\sigma^2_{1|2})
%	\frac{1+\big[
%	He_{4,0}(y_1,y_2) + He_{3,1}(y_1,y_2) + He_{2,2}(y_1,y_2)
%	\big]
%	}
%	{1+He_4(y_2)}
%\end{equation} 

%It does not look promising to the present author that one can demonstrate the marginal property and compute the conditional statistics, such as $\mathbb E[z_1|z_2]$, directly through above distribution. 
We can directly apply Eq.~(\ref{pred_mean}) to obtain the conditional mean in the simple example where the tilded indices there only apply to $z_2$. Surprisingly, the conditional mean for such noiseless case is
\begin{equation}
	\mathbb E[z_1|z_2]=%\frac{cz_2+\frac{cz_2H_4(z_2)}{16N}}
	%{1+\frac{H_4(z_2)}{16N}}
	cz_2\:,
\end{equation} coinciding with that in the GP limit because the {\it accidental} cancelation in the third-order tensor $A_{222}\propto(cV_{2222}-V_{1222})$. We shall show that the cancelation does not occur in the noisy (next Section) and more general cases. Application of Eq.~(\ref{pred_2nd_moment}) together with the above conditional mean results in the following conditional variance, 
\begin{equation}
	{\rm Var}[z_1|z_2]=%\mathbb E[z_1^2|z_2] - (\mathbb E[z_1|z_2])^2=
	(1-c^2)\bigg[1+
	\frac{4(2-c^2)}
	{16N+H_4(z_2)}H_2(z_2)
	\bigg]\:,
\end{equation} which consists of the corresponding variance in GP limit along with a term depending on $z^2_2$.

%On the other hand, the differential representation of joint distribution By applying Eq.~(\ref{pred_mean}) and (\ref{pred_2nd_moment}), one can show that

%By definition, the conditional distribution is obtained from dividing the joint distribution by the marginal one, 
%\begin{equation}
	%\int dz_{\alpha_1}q(z_{\alpha_1},\cdots,z_{\alpha_M}) = q(z_{\alpha_2},\cdots,z_{\alpha_M})
%	q(z_1|\tilde{\bf z})=\frac{q(z_1,\tilde{\bf z})}{q(\tilde{\bf z})}\:.
%\end{equation} It follows from the Edgeworth series, the conditional distribution up to $O(1/N)$
%\begin{equation}
%\begin{split}
%	q(z_1|\tilde{\bf z}) &= \frac{
%	e^{-\frac{(z_1-\mu)^2}{2\sigma^2}}
%	}{\sqrt{2\pi\sigma^2}}
%	\bigg[
%	1 + \frac{V}{24N}(\chi^{(0)} -\chi^{(1)}_{ij}z_iz_j+\chi^{(2)}_{ijlm}z_iz_jz_lz_m) \\
%	&-\frac{\tilde V}{24N}(
%	\tilde\chi^{(0)} -\tilde\chi^{(1)}_{i'j'}z_{i'}z_{j'}+\tilde\chi^{(2)}_{i'j'l'm'}z_{i'}z_{j'}z_{l'}z_{m'}
%	)
%	\bigg]
%\end{split}
%\end{equation} In order to have the conditional distribution normalized, the candidate form 
%\begin{equation}
%	q(z_1|\tilde{\bf z}) = \frac{e^{-\frac{(z_1-\mu_a)^2}{2\sigma_a^2}}}{\sqrt{2\pi\sigma_a^2}}\big\{
%	1+\frac{V_a}{24N}[3-6(\frac{z_1-\mu_a}{\sigma_a})^2+(\frac{z_1-\mu_a}{\sigma_a})^4]
%	\big\}
%\end{equation} with the undetermined coefficients: the adjusted conditional mean $\mu_a$, conditional variance $\sigma_a^2$, and the fourth order cumulant $V_a$.

\section{Bayesian Regression with non-gaussian prior}\label{Bayesian}
Having established the marginal and conditional properties of the non-Gaussian distribution in Eq.~(\ref{Edgeworth_Joint}), we now continue investigating the posterior distribution over the unseen function value $z_*$ associated with input ${\bf x}_*$ when the noisy observations $\mathcal D=\{({\bf x}_1,y_1),({\bf x}_2,y_2),\cdots\}$ are known. 
%Here, we briefly review the case for Gaussian distribution, which corresponds to the infinitely wide network. 
According to Bayes's rule, the objective distribution is the posterior distribution defined as,
\begin{equation}
	q(z_*|\mathcal D)=\frac{q(z_*,\mathcal D)}{q(\mathcal D)}=\frac{\int d{\bf z}\ q(z_*,{\bf z})\mathcal N({\bf y}|{\bf z},\Lambda)}
	{\int d{\bf z}\ q({\bf z})\mathcal N({\bf y}|{\bf z},\Lambda)}\:,
\end{equation} with $\Lambda$ denoting a diagonal matrix representing the noise in Gaussian likelihood. When the $q$'s in the numerator and denominator are both Gaussian, which corresponds to the infinitely wide network, the limiting distribution reads
\begin{equation*}
\begin{split}
	q_{\infty}(z_*|\mathcal D) &= \int d{\bf z}\ \mathcal N(z_*|{\bf k}_*^tK^{-1}{\bf z},k_{**}-{\bf k}_*^tK^{-1}{\bf k}_*)
	\ \mathcal N({\bf z}|K(K+\Lambda)^{-1}{\bf y}, \Lambda K(K+\Lambda)^{-1})\\
	&= \mathcal N\big(
	z_*|{\bf k}_*^t(K+\Lambda)^{-1}{\bf y}, k_{**}-{\bf k}_*^t(K+\Lambda)^{-1}{\bf k}_*
	\big)\:.
\end{split}
\end{equation*} In above first equality, the first and second Gaussian distributions on the right hand side represent the conditional density $q_{\infty}(z_*|{\bf z})$ and the posterior $q_{\infty}({\bf z}|\mathcal D)$, respectively.

%\subsection{Regression}
As for the case of finite-width, the involved Gaussian likelihood can be dealt with by continuing application of the properties of adjoint differential operator as well as derivative of independent Gaussians. The details of derivation of evidence $q(\mathcal D)$ can be seen in Appendix~\ref{evidence_proof}, and the expression in terms of derivatives with respect to the observed $y$'s reads,
\begin{equation}
\begin{split}
	q(\mathcal D)&=\int d{\bf z}\ \mathcal N({\bf y}|{\bf z},\Lambda) q({\bf z})\\
	&=(1+v_{ijml}\partial_{y_i}\partial_{y_j}\partial_{y_m}\partial_{y_l})
	\mathcal N({\bf y}|0,K+\Lambda)\:.\label{evidence}
\end{split}
\end{equation}
Despite its simple form, the evidence term can not be further be manipulated like in the infinite-width case. The details of derivation of the posterior distribution for output $z_*$ at the unseen input, %is given by,
\begin{equation}
\begin{split}
	q(z_*|\mathcal D) &= \frac{
	\int d{\bf z}\ \mathcal N({\bf y}|{\bf z},\Lambda)q(z_*,{\bf z})
	}
	{
	%\int d{\bf z}\ \mathcal N({\bf y}|{\bf z},\Lambda)q({\bf z})
	q(\mathcal D)
	}\\
	&=
	(1+v_{\hat i\hat j\hat m\hat l}\partial_{\hat i\hat j\hat m\hat l})
	\mathcal N\big[\bigl(\begin{smallmatrix}z_* \\{\bf y} \end{smallmatrix}\bigr)
	\big|0,\bigl(
	\begin{smallmatrix}1 &{\bf k}_*^t\\{\bf k}_*&K+\Lambda \end{smallmatrix}\bigr)
	%z_*|{\bf k}^t_*(K+\Lambda)^{-1}{\bf y},k_{**}-{\bf k}^t_*(K+\Lambda)^{-1}{\bf k}_*
	\big]
	%\bigg[
	%\mathcal N\big(
	%z_*|{\bf k}^t_*(K+\Lambda)^{-1}{\bf y},k_{**}-{\bf k}^t_*(K+\Lambda)^{-1}{\bf k}_*
	%\big)
	%\mathcal N({\bf y}|0,K+\Lambda)
	%\bigg]
	/q(\mathcal D)\:,\label{bayes_posterior}
\end{split}
\end{equation} can also be found in Appendix~\ref{evidence_proof}. Here, the notations can be understood as follows: the hatted indices, $\hat i$ for instance, additionally include the symbol $*$ associated with test function value $z_*$, and the derivative refers to $\partial_*:=\partial_{z_*}$ (with respect to function value) and $\partial_i:=\partial_{y_i}$ (with respect to the observed value).

%to demonstrate an application of above formulation. The shallow and finitely wide network defined in Eq.~(\ref{shallow_network}) with scalar output is the model. Given a single observation $y$ at input ${\bf x}$, what is the distribution $p(y_*|y)$ over the unobserved $y_*$ associated with the input at ${\bf x}_*$? If one directly applies the expressions of GP inference and neglect the noise during observation, the resultant distribution is a Gaussian with predictive mean $\mu=cy$ and variance $\sigma^2=1-c^2$. Here we use $c$ to denote the covariance $k({\bf x, x}_*)=\exp(|{\bf x- x}_*|^2/2d)$ defined in Eq.(\ref{2nd_moment}). Next, we shall see the effect of finite $N$ on the prediction. Note that the predictive distribution is non-Gaussian. Nevertheless, we can still compute the predictive mean and variance by exploiting Eq.~(\ref{pred_mean}) and (\ref{pred_2nd_moment}).

We conclude this Section with considering the same simple regression example but with noisy observation in $y_2$ at input ${\bf x}_2$, and we wish to predict the unobserved function value $z_1$. To stress the role of noise parameter $\sigma_n$, we only show the predictive mean here, %In the simple bivariate case where 
\begin{equation}
	\mathbb E[z_1|y_2]=c\alpha_n^2y_2-%\frac{cy_2}{1+\sigma_n^2}-
	\frac{c\sigma_n^2\alpha_n^5H_3(\alpha_ny_2)}{4N+\alpha_n^4H_4(\alpha_ny_2)}\:,%\frac{y_2}{\sqrt{1+\sigma_n^2}})
\end{equation} with $\alpha_n=1/\sqrt{1+\sigma_n^2}$. The correction term vanishes as the noise parameter does, and is odd with respect to the sign change of $y_2$.

\section{Deep Bottlenecked Network}\label{bottleneck}

Up to this point, we have focused on the emergence of non-Gaussianity in prior function distribution in wide and shallow network, as well as how the non-Gaussian prior affects the inference in a regression task. Our approach can be extended to the deep and wide neural networks as the corresponding prior distributions approach GP~\cite{matthews2018gaussian,lee2018deep}. However, such wide-width assumption leading to {\it weak} non-Gaussianity is not valid for the bottlenecked networks in which the hidden layers are alternatively wide and narrow~\cite{cutajar2017random,agrawal2020wide}.
%\subsection{Classification}

%\subsection{A simple (but nontrivial) application}
Let us consider a two-layer bottlenecked feed forward network modeling the hierarchical mapping ${\bf x}\in\mathbb R^d\mapsto {\bf z}^{(1)}\in\mathbb R^H\mapsto z^{(2)}\in\mathbb R$. The hierarchy consists of the following computations,%~\cite{cutajar2017random,agrawal2020wide},
\begin{equation}
\begin{split}
	z^{(1)}_{i,\alpha}&=\sqrt{\frac{2}{N_1}}\sum_{j=1}^{N_1}w^{(1)}_{ij}\cos(
	\frac{\Omega^{(1)}_j\cdot{\bf x}_{\alpha}}{\sqrt{d}}
	+\phi^{(1)}_j)\:,\\
	z^{(2)}_{\alpha}&=\sqrt{\frac{2}{N_2}}\sum_{j=1}^{N_2}w^{(2)}_{j}\cos(
	\frac{\Omega^{(2)}_j\cdot{\bf z}^{(1)}_{\alpha}}{\sqrt{H}}
	+\phi^{(2)}_j)\:,\label{2layer_network}
\end{split}
\end{equation} where the parameters $w$'s, $\Omega$'s, and $\phi$'s share the same prior distribution with their counterparts in the shallow network. The widths of hidden layers, $N_1$ and $N_2$, are large but the hidden output dimension $H$ is not. In the limit $N_{1,2}\rightarrow\infty$ while $H$ remains finite, the two-layer cosine bottlenecked network corresponds to the prior of deep Gaussian process~\cite{damianou2013deep,dunlop2018deep,pleiss2021limitations} with Gaussian kernel, which is a flexible and expressive function prior due to its compositional nature. As its name suggests, the conditional prior $z^{(2)}|{\bf z}^{(1)}$ is a GP and so is each component $z_i^{(1)}|{\bf x}$ in the hidden output. However, the marginal distribution for $z^{(2)}|{\bf x}$ is non-Gaussian as~\cite{lu2020interpretable} showed that the fourth cumulant is positive, i.e., a heavy-tailed distribution~\cite{vladimirova2019understanding}.

Here, we are interested in tracking how the small parameters, $1/N_1$ and $1/N_2$, and the bottleneck parameter $1/H$ enters the second moment. For a deep and finitely wide linear network, the prior is non-Gaussian~\cite{yaida2020non,zavatone2021exact} but the second moment does not receive correction due to the finite width~\cite{yaida2020non,roberts2021principles}. Thus, the effects of nonlinear activation on higher statistical moments are interesting (also see a recent work in~\cite{zavatone2021asymptotics}). For the deep cosine network in Eq.~(\ref{2layer_network}), we can first consider the wide limit, $N_{1,2}\rightarrow\infty$, but the bottleneck width $H$ remains finite. Following the result in Eq.~(\ref{2nd_moment}), the covariance for the deep model can be computed as followings, 
\begin{equation}
\begin{split}
	k^{(2)}_{\infty}({\bf x}_{\alpha},{\bf x}_{\beta})&=
	%\mathbb E[e^{-\frac{1}{2H}|{\bf z}_{\alpha}-{\bf z}_{\beta}|^2}] =
	%\mathbb \prod_{i=1}^H \mathbb E[e^{-\frac{1}{2H}(z_{i,\alpha}-z_{i,\beta})^2}]
	\mathbb E_{q_{\infty}}\bigg\{\exp\big(
	-\frac{|{\bf z}^{(1)}_{\alpha}-{\bf z}^{(1)}_{\beta}|^2}{2H}
	\big)\bigg\}\\
	&= \bigg[\int dz_{\alpha}dz_{\beta}\ e^{-\frac{(z_{\alpha}-z_{\beta})^2}{2H}} 
	q_{\infty}(z_{\alpha},z_{\beta})\label{dgp_2nd_moment}\bigg]^H\\
	&=\big[1+\frac{1-\exp(-\frac{|{\bf x}_{\alpha}-{\bf x}_{\beta}|^2}{2d})}{H/2}\big]^{-H/2}\:.
\end{split}
\end{equation} Note that the decomposition $\mathcal N(z_1,z_2|0,K)=\mathcal N(z_1+z_2|0,\sigma_+^2)\mathcal N(z_1-z_2|0,\sigma_-^2)$~\cite{lu2018standing} has been used in derivation with the variances $\sigma_{\pm}^2=2(K_{11}\pm K_{12})$. We want to stress that the above kernel is exact for any $H$. For a very narrow case, i.e., $H=1$, the above result coincides with that in~\cite{lu2020interpretable}, while in the very wide bottleneck case, $H\gg1$, it can be shown that 
\begin{equation}
	k^{(2)}_{\infty}({\bf x}_{\alpha},{\bf x}_{\beta})\approx\exp[k({\bf x}_{\alpha},{\bf x}_{\beta})-1]
	\big\{1+\frac{1}{H}[k({\bf x}_{\alpha},{\bf x}_{\beta})-1]^2\big\}\:,\label{limiting_kernel}
\end{equation} where the covariance $k$ in shallow network is given in (\ref{2nd_moment}). In addition, the exponential of shallow kernel in above corresponds to $H\rightarrow\infty$, which %the term in the first square bracket, 
is the same as the deep kernel in~\cite{duvenaud2014avoiding}. Indeed, as suggested in~\cite{roberts2021principles}, the appearance of reciprocal width, $1/H$, in the correction term is in association with the weak non-Gaussianity even though the other widths $N_{1,2}$ are infinite. %However, we remark that for intermediate width $H$ 

Next, we shall compute the covariance for the case where all widths are finite. In fact, the outer width $N_2$ does not enter the kernel and we only need large but finite $N_1$. The computation is similar and one only needs to replace the limiting joint distribution $q_{\infty}(z^{(1)}_{\alpha},z^{(1)}_{\beta})$ with the non-Gaussian $q$. The details of derivation using the property derivative of Gaussian can be seen in Appendix~\ref{2nd_moment_correction}. The kernel of the two-layer bottlenecked cosine network with $N_{1,2}<\infty$ reads,
\begin{equation}
	k^{(2)}({\bf x}_{\alpha},{\bf x}_{\beta})=\bigg[
	1+\frac{1-k({\bf x}_{\alpha},{\bf x}_{\beta})}{H/2}
	\bigg]^{-H/2}(1+\epsilon)^H\:,
\end{equation} with the correction due to finite $N_1$,
\begin{equation}
	\epsilon=\frac{V_{\alpha\alpha\alpha\alpha}+V_{\beta\beta\beta\beta}
	-4V_{\alpha\beta\beta\beta}-4V_{\beta\alpha\alpha\alpha}
	+6V_{\alpha\alpha\beta\beta}}{24}
	\big[\frac{3}{H^2}
	-\frac{6\sigma_-^2}{H^3(1+\sigma_-^2/H)}
	+\frac{3\sigma_-^4}{H^4(1+\sigma_-^2/H)^2}\big]\:,
\end{equation} with $\sigma_-^2=2(1-k)$ is used for easing the notation. Again, the above result is for general $H$. For the wide bottleneck limit, one can show that these small parameters enter the kernel with the correction of $O(\frac{1}{H})$ followed by $O(\frac{1}{N_1H})$. Hence, the deep model as a function prior serves as a GP with random kernel whose mean value converges to Eq.~(\ref{limiting_kernel}) when all widths $N_{1,2}$ and $H$ approach infinity. However, the role of $H$ is different from $N_1$ from our perturbative analysis, and $1/N_1$ does not appear alone in the small parameter. The influence of finite width on the kernel of deep and nonlinear and convolutional models is also studied in~\cite{aitchison2020bigger,li2022neural,hanin2022correlation}. %discussed the distribution of A preliminary discussion of the effect of nonlinear activation units in deep and convolutional networks can be found in~\cite{aitchison2020bigger}.  

The fourth moment can be computed in a similar manner. The closed form for $N_{1,2}\rightarrow\infty$ can be found in~\cite{lu2020interpretable}. For the finite width case, the fourth cumulant $V_{\alpha\beta\gamma\delta}^{(2)}$ becomes the $H$-th power of the permutational symmetrization of $\frac{1}{N_2}\mathbb E_q\{\exp[-(z^{(1)}_{\alpha}+z^{(1)}_{\beta}-z^{(1)}_{\gamma}-z^{(1)}_{\delta})^2/2H)\}$, the small parameters of which shall consist of $1/N_2$, $1/(N_2H)$, and $1/(N_2HN_1)$. The analysis of scaling with respect to the reciprocal width is more complex than the deep linear network.

\section{discussions}\label{discussion}
In essence, the finite width in random neural networks induces nonzero variance of kernel since the fourth cumulant is not zero. From function space perspective, the shallow networks with finitely large width can be regarded as a GP but the kernel itself is a random variable too, so the learned data representation is a distribution over kernels. Therefore, neural network with very narrow layers may not have enough expressive power for learning while the network with very wide layers is not flexible enough because it is equivalent to GP learning with one fixed kernel. It was recently suggested in Ref.~\onlinecite{ariosto2022statistical} through studying the partition function~\cite{seung1992statistical,li2021statistical} of finite deep network that Student-t process~\cite{shah2014student} may suitably represent finite-width network, which offers an alternative description than this paper. The reports in~\cite{aitchison2020bigger,pleiss2021limitations} observed the degradation of performance for Bayesian deep neural network when expanding the widths while the study in~\cite{lee2020finite} suggested otherwise.

Taking the wide limit with the perturbative approach is appealing because of the analytic and %expressions as a result of the 
elegant expressions for inference, which may shed light on future investigation of alternative algorithm for classification~\cite{csato1999efficient}, study of average test error in regression~\cite{sollich1998learning,williams2000upper} and classification~\cite{seeger2002pac} tasks using finite-width networks. In this paper, the conditional statistics and perturbed posterior distribution over the unseen output using the non-Gaussian prior are obtained with help of the differential representation of multivariate Edgeworth expansion. In parallel with the equivalence between GP prior and the random wide neural network, the investigation of neural tangent kernel~\cite{jacot2018neural,arora2019exact,hanin2019finite} is important for understanding the learning dynamics during the optimization and the relation with Bayesian neural network learning~\cite{karakida2019universal,khan2019approximate}. 
%our formulation of Bayesian regression with the non-Gaussian prior still have analytic forms.      

%\cite{DBLP:journals/corr/abs-2106-00651}
%\cite{li2021statistical,ariosto2022statistical}

%\cite{gabrie2020mean}

\section*{acknowledgement}
The research is supported by the Dean Office of School of Arts and Science at Rutgers University Newark. Correspondences with Jacob Zavatone-Veth and Pietro Rotondo are acknowledged.

\bibliography{/Users/felix/manuscript/Reference/reference1}

\begin{thebibliography}{44}%
\makeatletter
\providecommand \@ifxundefined [1]{%
 \@ifx{#1\undefined}
}%
\providecommand \@ifnum [1]{%
 \ifnum #1\expandafter \@firstoftwo
 \else \expandafter \@secondoftwo
 \fi
}%
\providecommand \@ifx [1]{%
 \ifx #1\expandafter \@firstoftwo
 \else \expandafter \@secondoftwo
 \fi
}%
\providecommand \natexlab [1]{#1}%
\providecommand \enquote  [1]{``#1''}%
\providecommand \bibnamefont  [1]{#1}%
\providecommand \bibfnamefont [1]{#1}%
\providecommand \citenamefont [1]{#1}%
\providecommand \href@noop [0]{\@secondoftwo}%
\providecommand \href [0]{\begingroup \@sanitize@url \@href}%
\providecommand \@href[1]{\@@startlink{#1}\@@href}%
\providecommand \@@href[1]{\endgroup#1\@@endlink}%
\providecommand \@sanitize@url [0]{\catcode `\\12\catcode `\$12\catcode
  `\&12\catcode `\#12\catcode `\^12\catcode `\_12\catcode `\%12\relax}%
\providecommand \@@startlink[1]{}%
\providecommand \@@endlink[0]{}%
\providecommand \url  [0]{\begingroup\@sanitize@url \@url }%
\providecommand \@url [1]{\endgroup\@href {#1}{\urlprefix }}%
\providecommand \urlprefix  [0]{URL }%
\providecommand \Eprint [0]{\href }%
\providecommand \doibase [0]{https://doi.org/}%
\providecommand \selectlanguage [0]{\@gobble}%
\providecommand \bibinfo  [0]{\@secondoftwo}%
\providecommand \bibfield  [0]{\@secondoftwo}%
\providecommand \translation [1]{[#1]}%
\providecommand \BibitemOpen [0]{}%
\providecommand \bibitemStop [0]{}%
\providecommand \bibitemNoStop [0]{.\EOS\space}%
\providecommand \EOS [0]{\spacefactor3000\relax}%
\providecommand \BibitemShut  [1]{\csname bibitem#1\endcsname}%
\let\auto@bib@innerbib\@empty
%</preamble>
\bibitem [{\citenamefont {Neal}(1997)}]{neal1997monte}%
  \BibitemOpen
  \bibfield  {author} {\bibinfo {author} {\bibfnamefont {R.~M.}\ \bibnamefont
  {Neal}},\ }\bibfield  {title} {\bibinfo {title} {Monte carlo implementation
  of {Gaussian} process models for {Bayesian} regression and classification},\
  }\href@noop {} {\bibfield  {journal} {\bibinfo  {journal} {arXiv preprint
  physics/9701026}\ } (\bibinfo {year} {1997})}\BibitemShut {NoStop}%
\bibitem [{\citenamefont {Rasmussen}\ and\ \citenamefont
  {Williams}(2006)}]{rasmussen2006gaussian}%
  \BibitemOpen
  \bibfield  {author} {\bibinfo {author} {\bibfnamefont {C.~E.}\ \bibnamefont
  {Rasmussen}}\ and\ \bibinfo {author} {\bibfnamefont {C.~K.~I.}\ \bibnamefont
  {Williams}},\ }\href@noop {} {\emph {\bibinfo {title} {Gaussian Process for
  Machine Learning}}}\ (\bibinfo  {publisher} {MIT press},\ \bibinfo {address}
  {Cambridge, MA},\ \bibinfo {year} {2006})\BibitemShut {NoStop}%
\bibitem [{\citenamefont {Williams}(1997)}]{williams1997computing}%
  \BibitemOpen
  \bibfield  {author} {\bibinfo {author} {\bibfnamefont {C.~K.}\ \bibnamefont
  {Williams}},\ }\bibfield  {title} {\bibinfo {title} {Computing with infinite
  networks},\ }in\ \href@noop {} {\emph {\bibinfo {booktitle} {Advances in
  neural information processing systems}}}\ (\bibinfo {year} {1997})\ pp.\
  \bibinfo {pages} {295--301}\BibitemShut {NoStop}%
\bibitem [{\citenamefont {Cho}\ and\ \citenamefont
  {Saul}(2009)}]{cho2009kernel}%
  \BibitemOpen
  \bibfield  {author} {\bibinfo {author} {\bibfnamefont {Y.}~\bibnamefont
  {Cho}}\ and\ \bibinfo {author} {\bibfnamefont {L.~K.}\ \bibnamefont {Saul}},\
  }\bibfield  {title} {\bibinfo {title} {Kernel methods for deep learning},\
  }in\ \href@noop {} {\emph {\bibinfo {booktitle} {Advances in neural
  information processing systems}}}\ (\bibinfo {year} {2009})\ pp.\ \bibinfo
  {pages} {342--350}\BibitemShut {NoStop}%
\bibitem [{\citenamefont {Matthews}\ \emph {et~al.}(2018)\citenamefont
  {Matthews}, \citenamefont {Hron}, \citenamefont {Rowland}, \citenamefont
  {Turner},\ and\ \citenamefont {Ghahramani}}]{matthews2018gaussian}%
  \BibitemOpen
  \bibfield  {author} {\bibinfo {author} {\bibfnamefont {A.~G. d.~G.}\
  \bibnamefont {Matthews}}, \bibinfo {author} {\bibfnamefont {J.}~\bibnamefont
  {Hron}}, \bibinfo {author} {\bibfnamefont {M.}~\bibnamefont {Rowland}},
  \bibinfo {author} {\bibfnamefont {R.~E.}\ \bibnamefont {Turner}},\ and\
  \bibinfo {author} {\bibfnamefont {Z.}~\bibnamefont {Ghahramani}},\ }\bibfield
   {title} {\bibinfo {title} {Gaussian process behaviour in wide deep neural
  networks},\ }in\ \href@noop {} {\emph {\bibinfo {booktitle} {International
  Conference on Learning Representations}}}\ (\bibinfo {year}
  {2018})\BibitemShut {NoStop}%
\bibitem [{\citenamefont {Lee}\ \emph {et~al.}(2018)\citenamefont {Lee},
  \citenamefont {Bahri}, \citenamefont {Novak}, \citenamefont {Schoenholz},
  \citenamefont {Pennington},\ and\ \citenamefont
  {Sohl-Dickstein}}]{lee2018deep}%
  \BibitemOpen
  \bibfield  {author} {\bibinfo {author} {\bibfnamefont {J.}~\bibnamefont
  {Lee}}, \bibinfo {author} {\bibfnamefont {Y.}~\bibnamefont {Bahri}}, \bibinfo
  {author} {\bibfnamefont {R.}~\bibnamefont {Novak}}, \bibinfo {author}
  {\bibfnamefont {S.~S.}\ \bibnamefont {Schoenholz}}, \bibinfo {author}
  {\bibfnamefont {J.}~\bibnamefont {Pennington}},\ and\ \bibinfo {author}
  {\bibfnamefont {J.}~\bibnamefont {Sohl-Dickstein}},\ }\bibfield  {title}
  {\bibinfo {title} {Deep neural networks as {Gaussian} processes},\ }in\
  \href@noop {} {\emph {\bibinfo {booktitle} {International Conference on
  Learning Representations}}}\ (\bibinfo {year} {2018})\BibitemShut {NoStop}%
\bibitem [{\citenamefont {Osband}(2016)}]{osband2016risk}%
  \BibitemOpen
  \bibfield  {author} {\bibinfo {author} {\bibfnamefont {I.}~\bibnamefont
  {Osband}},\ }\bibfield  {title} {\bibinfo {title} {Risk versus uncertainty in
  deep learning: Bayes, bootstrap and the dangers of dropout},\ }in\ \href@noop
  {} {\emph {\bibinfo {booktitle} {Workshop on Bayesian Deep Learning}}}\
  (\bibinfo {organization} {NIPS},\ \bibinfo {year} {2016})\BibitemShut
  {NoStop}%
\bibitem [{\citenamefont {Yaida}(2020)}]{yaida2020non}%
  \BibitemOpen
  \bibfield  {author} {\bibinfo {author} {\bibfnamefont {S.}~\bibnamefont
  {Yaida}},\ }\bibfield  {title} {\bibinfo {title} {Non-{Gaussian} processes
  and neural networks at finite widths},\ }in\ \href@noop {} {\emph {\bibinfo
  {booktitle} {Mathematical and Scientific Machine Learning}}}\ (\bibinfo
  {organization} {PMLR},\ \bibinfo {year} {2020})\ pp.\ \bibinfo {pages}
  {165--192}\BibitemShut {NoStop}%
\bibitem [{\citenamefont {Roberts}\ \emph {et~al.}(2021)\citenamefont
  {Roberts}, \citenamefont {Yaida},\ and\ \citenamefont
  {Hanin}}]{roberts2021principles}%
  \BibitemOpen
  \bibfield  {author} {\bibinfo {author} {\bibfnamefont {D.~A.}\ \bibnamefont
  {Roberts}}, \bibinfo {author} {\bibfnamefont {S.}~\bibnamefont {Yaida}},\
  and\ \bibinfo {author} {\bibfnamefont {B.}~\bibnamefont {Hanin}},\ }\bibfield
   {title} {\bibinfo {title} {The principles of deep learning theory},\
  }\href@noop {} {\bibfield  {journal} {\bibinfo  {journal} {arXiv preprint
  arXiv:2106.10165}\ } (\bibinfo {year} {2021})}\BibitemShut {NoStop}%
\bibitem [{\citenamefont {Dyer}\ and\ \citenamefont
  {Gur-Ari}(2019)}]{dyer2019asymptotics}%
  \BibitemOpen
  \bibfield  {author} {\bibinfo {author} {\bibfnamefont {E.}~\bibnamefont
  {Dyer}}\ and\ \bibinfo {author} {\bibfnamefont {G.}~\bibnamefont {Gur-Ari}},\
  }\bibfield  {title} {\bibinfo {title} {Asymptotics of wide networks from
  {Feynman} diagrams},\ }in\ \href@noop {} {\emph {\bibinfo {booktitle}
  {International Conference on Learning Representations}}}\ (\bibinfo {year}
  {2019})\BibitemShut {NoStop}%
\bibitem [{\citenamefont {Gabri{\'e}}(2020)}]{gabrie2020mean}%
  \BibitemOpen
  \bibfield  {author} {\bibinfo {author} {\bibfnamefont {M.}~\bibnamefont
  {Gabri{\'e}}},\ }\bibfield  {title} {\bibinfo {title} {Mean-field inference
  methods for neural networks},\ }\href@noop {} {\bibfield  {journal} {\bibinfo
   {journal} {Journal of Physics A: Mathematical and Theoretical}\ }\textbf
  {\bibinfo {volume} {53}},\ \bibinfo {pages} {223002} (\bibinfo {year}
  {2020})}\BibitemShut {NoStop}%
\bibitem [{\citenamefont {Skovgaard}(1986)}]{skovgaard1986multivariate}%
  \BibitemOpen
  \bibfield  {author} {\bibinfo {author} {\bibfnamefont {I.~M.}\ \bibnamefont
  {Skovgaard}},\ }\bibfield  {title} {\bibinfo {title} {On multivariate
  {Edgeworth} expansions},\ }\href@noop {} {\bibfield  {journal} {\bibinfo
  {journal} {International Statistical Review/Revue Internationale de
  Statistique}\ ,\ \bibinfo {pages} {169}} (\bibinfo {year}
  {1986})}\BibitemShut {NoStop}%
\bibitem [{\citenamefont {Lu}\ \emph {et~al.}(2020)\citenamefont {Lu},
  \citenamefont {Yang}, \citenamefont {Hao},\ and\ \citenamefont
  {Shafto}}]{lu2020interpretable}%
  \BibitemOpen
  \bibfield  {author} {\bibinfo {author} {\bibfnamefont {C.-K.}\ \bibnamefont
  {Lu}}, \bibinfo {author} {\bibfnamefont {S.~C.-H.}\ \bibnamefont {Yang}},
  \bibinfo {author} {\bibfnamefont {X.}~\bibnamefont {Hao}},\ and\ \bibinfo
  {author} {\bibfnamefont {P.}~\bibnamefont {Shafto}},\ }\bibfield  {title}
  {\bibinfo {title} {Interpretable deep {Gaussian} processes with moments},\
  }in\ \href@noop {} {\emph {\bibinfo {booktitle} {International Conference on
  Artificial Intelligence and Statistics}}}\ (\bibinfo {year} {2020})\ pp.\
  \bibinfo {pages} {613--623}\BibitemShut {NoStop}%
\bibitem [{\citenamefont {Agrawal}\ \emph {et~al.}(2020)\citenamefont
  {Agrawal}, \citenamefont {Papamarkou},\ and\ \citenamefont
  {Hinkle}}]{agrawal2020wide}%
  \BibitemOpen
  \bibfield  {author} {\bibinfo {author} {\bibfnamefont {D.}~\bibnamefont
  {Agrawal}}, \bibinfo {author} {\bibfnamefont {T.}~\bibnamefont
  {Papamarkou}},\ and\ \bibinfo {author} {\bibfnamefont {J.~D.}\ \bibnamefont
  {Hinkle}},\ }\bibfield  {title} {\bibinfo {title} {Wide neural networks with
  bottlenecks are deep {Gaussian} processes.},\ }\href@noop {} {\bibfield
  {journal} {\bibinfo  {journal} {J. Mach. Learn. Res.}\ }\textbf {\bibinfo
  {volume} {21}},\ \bibinfo {pages} {175} (\bibinfo {year} {2020})}\BibitemShut
  {NoStop}%
\bibitem [{\citenamefont {Damianou}\ and\ \citenamefont
  {Lawrence}(2013)}]{damianou2013deep}%
  \BibitemOpen
  \bibfield  {author} {\bibinfo {author} {\bibfnamefont {A.}~\bibnamefont
  {Damianou}}\ and\ \bibinfo {author} {\bibfnamefont {N.}~\bibnamefont
  {Lawrence}},\ }\bibfield  {title} {\bibinfo {title} {Deep {Gaussian}
  processes},\ }in\ \href@noop {} {\emph {\bibinfo {booktitle} {Artificial
  Intelligence and Statistics}}}\ (\bibinfo {year} {2013})\ pp.\ \bibinfo
  {pages} {207--215}\BibitemShut {NoStop}%
\bibitem [{\citenamefont {Rahimi}\ and\ \citenamefont
  {Recht}(2008)}]{rahimi2008random}%
  \BibitemOpen
  \bibfield  {author} {\bibinfo {author} {\bibfnamefont {A.}~\bibnamefont
  {Rahimi}}\ and\ \bibinfo {author} {\bibfnamefont {B.}~\bibnamefont {Recht}},\
  }\bibfield  {title} {\bibinfo {title} {Random features for large-scale kernel
  machines},\ }in\ \href@noop {} {\emph {\bibinfo {booktitle} {Advances in
  neural information processing systems}}}\ (\bibinfo {year} {2008})\ pp.\
  \bibinfo {pages} {1177--1184}\BibitemShut {NoStop}%
\bibitem [{\citenamefont {Jacot}\ \emph {et~al.}(2018)\citenamefont {Jacot},
  \citenamefont {Gabriel},\ and\ \citenamefont {Hongler}}]{jacot2018neural}%
  \BibitemOpen
  \bibfield  {author} {\bibinfo {author} {\bibfnamefont {A.}~\bibnamefont
  {Jacot}}, \bibinfo {author} {\bibfnamefont {F.}~\bibnamefont {Gabriel}},\
  and\ \bibinfo {author} {\bibfnamefont {C.}~\bibnamefont {Hongler}},\
  }\bibfield  {title} {\bibinfo {title} {Neural tangent kernel: Convergence and
  generalization in neural networks},\ }in\ \href@noop {} {\emph {\bibinfo
  {booktitle} {Advances in neural information processing systems}}}\ (\bibinfo
  {year} {2018})\ pp.\ \bibinfo {pages} {8571--8580}\BibitemShut {NoStop}%
\bibitem [{\citenamefont {Antognini}(2019)}]{antognini2019finite}%
  \BibitemOpen
  \bibfield  {author} {\bibinfo {author} {\bibfnamefont {J.~M.}\ \bibnamefont
  {Antognini}},\ }\bibfield  {title} {\bibinfo {title} {Finite size corrections
  for neural network {Gaussian} processes},\ }\href@noop {} {\bibfield
  {journal} {\bibinfo  {journal} {arXiv preprint arXiv:1908.10030}\ } (\bibinfo
  {year} {2019})}\BibitemShut {NoStop}%
\bibitem [{\citenamefont {Papoulis}\ and\ \citenamefont
  {Unnikrishna~Pillai}(2002)}]{papoulis2002probability}%
  \BibitemOpen
  \bibfield  {author} {\bibinfo {author} {\bibfnamefont {A.}~\bibnamefont
  {Papoulis}}\ and\ \bibinfo {author} {\bibfnamefont {S.}~\bibnamefont
  {Unnikrishna~Pillai}},\ }\href@noop {} {\emph {\bibinfo {title} {Probability,
  random variables and stochastic processes}}}\ (\bibinfo {year}
  {2002})\BibitemShut {NoStop}%
\bibitem [{\citenamefont {Welling}(2005)}]{welling2005robust}%
  \BibitemOpen
  \bibfield  {author} {\bibinfo {author} {\bibfnamefont {M.}~\bibnamefont
  {Welling}},\ }\bibfield  {title} {\bibinfo {title} {Robust higher order
  statistics},\ }in\ \href@noop {} {\emph {\bibinfo {booktitle} {International
  Workshop on Artificial Intelligence and Statistics}}}\ (\bibinfo
  {organization} {PMLR},\ \bibinfo {year} {2005})\ pp.\ \bibinfo {pages}
  {405--412}\BibitemShut {NoStop}%
\bibitem [{\citenamefont {Shah}\ \emph {et~al.}(2014)\citenamefont {Shah},
  \citenamefont {Wilson},\ and\ \citenamefont {Ghahramani}}]{shah2014student}%
  \BibitemOpen
  \bibfield  {author} {\bibinfo {author} {\bibfnamefont {A.}~\bibnamefont
  {Shah}}, \bibinfo {author} {\bibfnamefont {A.}~\bibnamefont {Wilson}},\ and\
  \bibinfo {author} {\bibfnamefont {Z.}~\bibnamefont {Ghahramani}},\ }\bibfield
   {title} {\bibinfo {title} {Student-t processes as alternatives to {Gaussian}
  processes},\ }in\ \href@noop {} {\emph {\bibinfo {booktitle} {Artificial
  intelligence and statistics}}}\ (\bibinfo {year} {2014})\ pp.\ \bibinfo
  {pages} {877--885}\BibitemShut {NoStop}%
\bibitem [{\citenamefont {Cutajar}\ \emph {et~al.}(2017)\citenamefont
  {Cutajar}, \citenamefont {Bonilla}, \citenamefont {Michiardi},\ and\
  \citenamefont {Filippone}}]{cutajar2017random}%
  \BibitemOpen
  \bibfield  {author} {\bibinfo {author} {\bibfnamefont {K.}~\bibnamefont
  {Cutajar}}, \bibinfo {author} {\bibfnamefont {E.~V.}\ \bibnamefont
  {Bonilla}}, \bibinfo {author} {\bibfnamefont {P.}~\bibnamefont {Michiardi}},\
  and\ \bibinfo {author} {\bibfnamefont {M.}~\bibnamefont {Filippone}},\
  }\bibfield  {title} {\bibinfo {title} {Random feature expansions for deep
  {Gaussian} processes},\ }in\ \href@noop {} {\emph {\bibinfo {booktitle}
  {Proceedings of the 34th International Conference on Machine Learning-Volume
  70}}}\ (\bibinfo {organization} {JMLR. org},\ \bibinfo {year} {2017})\ pp.\
  \bibinfo {pages} {884--893}\BibitemShut {NoStop}%
\bibitem [{\citenamefont {Dunlop}\ \emph {et~al.}(2018)\citenamefont {Dunlop},
  \citenamefont {Girolami}, \citenamefont {Stuart},\ and\ \citenamefont
  {Teckentrup}}]{dunlop2018deep}%
  \BibitemOpen
  \bibfield  {author} {\bibinfo {author} {\bibfnamefont {M.~M.}\ \bibnamefont
  {Dunlop}}, \bibinfo {author} {\bibfnamefont {M.~A.}\ \bibnamefont
  {Girolami}}, \bibinfo {author} {\bibfnamefont {A.~M.}\ \bibnamefont
  {Stuart}},\ and\ \bibinfo {author} {\bibfnamefont {A.~L.}\ \bibnamefont
  {Teckentrup}},\ }\bibfield  {title} {\bibinfo {title} {How deep are deep
  {Gaussian} processes?},\ }\href@noop {} {\bibfield  {journal} {\bibinfo
  {journal} {The Journal of Machine Learning Research}\ }\textbf {\bibinfo
  {volume} {19}},\ \bibinfo {pages} {2100} (\bibinfo {year}
  {2018})}\BibitemShut {NoStop}%
\bibitem [{\citenamefont {Pleiss}\ and\ \citenamefont
  {Cunningham}(2021)}]{pleiss2021limitations}%
  \BibitemOpen
  \bibfield  {author} {\bibinfo {author} {\bibfnamefont {G.}~\bibnamefont
  {Pleiss}}\ and\ \bibinfo {author} {\bibfnamefont {J.~P.}\ \bibnamefont
  {Cunningham}},\ }\bibfield  {title} {\bibinfo {title} {The limitations of
  large width in neural networks: A deep {Gaussian} process perspective},\
  }\href@noop {} {\bibfield  {journal} {\bibinfo  {journal} {Advances in Neural
  Information Processing Systems}\ }\textbf {\bibinfo {volume} {34}} (\bibinfo
  {year} {2021})}\BibitemShut {NoStop}%
\bibitem [{\citenamefont {Vladimirova}\ \emph {et~al.}(2019)\citenamefont
  {Vladimirova}, \citenamefont {Verbeek}, \citenamefont {Mesejo},\ and\
  \citenamefont {Arbel}}]{vladimirova2019understanding}%
  \BibitemOpen
  \bibfield  {author} {\bibinfo {author} {\bibfnamefont {M.}~\bibnamefont
  {Vladimirova}}, \bibinfo {author} {\bibfnamefont {J.}~\bibnamefont
  {Verbeek}}, \bibinfo {author} {\bibfnamefont {P.}~\bibnamefont {Mesejo}},\
  and\ \bibinfo {author} {\bibfnamefont {J.}~\bibnamefont {Arbel}},\ }\bibfield
   {title} {\bibinfo {title} {Understanding priors in {Bayesian} neural
  networks at the unit level},\ }in\ \href@noop {} {\emph {\bibinfo {booktitle}
  {International Conference on Machine Learning}}}\ (\bibinfo {organization}
  {PMLR},\ \bibinfo {year} {2019})\ pp.\ \bibinfo {pages}
  {6458--6467}\BibitemShut {NoStop}%
\bibitem [{\citenamefont {Zavatone-Veth}\ and\ \citenamefont
  {Pehlevan}(2021)}]{zavatone2021exact}%
  \BibitemOpen
  \bibfield  {author} {\bibinfo {author} {\bibfnamefont {J.}~\bibnamefont
  {Zavatone-Veth}}\ and\ \bibinfo {author} {\bibfnamefont {C.}~\bibnamefont
  {Pehlevan}},\ }\bibfield  {title} {\bibinfo {title} {Exact marginal prior
  distributions of finite {Bayesian} neural networks},\ }\href@noop {}
  {\bibfield  {journal} {\bibinfo  {journal} {Advances in Neural Information
  Processing Systems}\ }\textbf {\bibinfo {volume} {34}} (\bibinfo {year}
  {2021})}\BibitemShut {NoStop}%
\bibitem [{\citenamefont {Zavatone-Veth}\ \emph {et~al.}(2021)\citenamefont
  {Zavatone-Veth}, \citenamefont {Canatar}, \citenamefont {Ruben},\ and\
  \citenamefont {Pehlevan}}]{zavatone2021asymptotics}%
  \BibitemOpen
  \bibfield  {author} {\bibinfo {author} {\bibfnamefont {J.}~\bibnamefont
  {Zavatone-Veth}}, \bibinfo {author} {\bibfnamefont {A.}~\bibnamefont
  {Canatar}}, \bibinfo {author} {\bibfnamefont {B.}~\bibnamefont {Ruben}},\
  and\ \bibinfo {author} {\bibfnamefont {C.}~\bibnamefont {Pehlevan}},\
  }\bibfield  {title} {\bibinfo {title} {Asymptotics of representation learning
  in finite bayesian neural networks},\ }\href@noop {} {\bibfield  {journal}
  {\bibinfo  {journal} {Advances in neural information processing systems}\
  }\textbf {\bibinfo {volume} {34}},\ \bibinfo {pages} {24765} (\bibinfo {year}
  {2021})}\BibitemShut {NoStop}%
\bibitem [{\citenamefont {Lu}\ \emph {et~al.}(2018)\citenamefont {Lu},
  \citenamefont {Yang},\ and\ \citenamefont {Shafto}}]{lu2018standing}%
  \BibitemOpen
  \bibfield  {author} {\bibinfo {author} {\bibfnamefont {C.-K.}\ \bibnamefont
  {Lu}}, \bibinfo {author} {\bibfnamefont {S.~C.-H.}\ \bibnamefont {Yang}},\
  and\ \bibinfo {author} {\bibfnamefont {P.}~\bibnamefont {Shafto}},\
  }\bibfield  {title} {\bibinfo {title} {Standing-wave-decomposition {Gaussian}
  process},\ }\href@noop {} {\bibfield  {journal} {\bibinfo  {journal}
  {Physical Review E}\ }\textbf {\bibinfo {volume} {98}},\ \bibinfo {pages}
  {032303} (\bibinfo {year} {2018})}\BibitemShut {NoStop}%
\bibitem [{\citenamefont {Duvenaud}\ \emph {et~al.}(2014)\citenamefont
  {Duvenaud}, \citenamefont {Rippel}, \citenamefont {Adams},\ and\
  \citenamefont {Ghahramani}}]{duvenaud2014avoiding}%
  \BibitemOpen
  \bibfield  {author} {\bibinfo {author} {\bibfnamefont {D.}~\bibnamefont
  {Duvenaud}}, \bibinfo {author} {\bibfnamefont {O.}~\bibnamefont {Rippel}},
  \bibinfo {author} {\bibfnamefont {R.}~\bibnamefont {Adams}},\ and\ \bibinfo
  {author} {\bibfnamefont {Z.}~\bibnamefont {Ghahramani}},\ }\bibfield  {title}
  {\bibinfo {title} {Avoiding pathologies in very deep networks},\ }in\
  \href@noop {} {\emph {\bibinfo {booktitle} {Artificial Intelligence and
  Statistics}}}\ (\bibinfo {year} {2014})\ pp.\ \bibinfo {pages}
  {202--210}\BibitemShut {NoStop}%
\bibitem [{\citenamefont {Aitchison}(2020)}]{aitchison2020bigger}%
  \BibitemOpen
  \bibfield  {author} {\bibinfo {author} {\bibfnamefont {L.}~\bibnamefont
  {Aitchison}},\ }\bibfield  {title} {\bibinfo {title} {Why bigger is not
  always better: on finite and infinite neural networks},\ }in\ \href@noop {}
  {\emph {\bibinfo {booktitle} {International Conference on Machine
  Learning}}}\ (\bibinfo {organization} {PMLR},\ \bibinfo {year} {2020})\ pp.\
  \bibinfo {pages} {156--164}\BibitemShut {NoStop}%
\bibitem [{\citenamefont {Li}\ \emph {et~al.}(2022)\citenamefont {Li},
  \citenamefont {Nica},\ and\ \citenamefont {Roy}}]{li2022neural}%
  \BibitemOpen
  \bibfield  {author} {\bibinfo {author} {\bibfnamefont {M.~B.}\ \bibnamefont
  {Li}}, \bibinfo {author} {\bibfnamefont {M.}~\bibnamefont {Nica}},\ and\
  \bibinfo {author} {\bibfnamefont {D.~M.}\ \bibnamefont {Roy}},\ }\bibfield
  {title} {\bibinfo {title} {The neural covariance sde: Shaped infinite
  depth-and-width networks at initialization},\ }\href@noop {} {\bibfield
  {journal} {\bibinfo  {journal} {arXiv preprint arXiv:2206.02768}\ } (\bibinfo
  {year} {2022})}\BibitemShut {NoStop}%
\bibitem [{\citenamefont {Hanin}(2022)}]{hanin2022correlation}%
  \BibitemOpen
  \bibfield  {author} {\bibinfo {author} {\bibfnamefont {B.}~\bibnamefont
  {Hanin}},\ }\bibfield  {title} {\bibinfo {title} {Correlation functions in
  random fully connected neural networks at finite width},\ }\href@noop {}
  {\bibfield  {journal} {\bibinfo  {journal} {arXiv preprint arXiv:2204.01058}\
  } (\bibinfo {year} {2022})}\BibitemShut {NoStop}%
\bibitem [{\citenamefont {Ariosto}\ \emph {et~al.}(2022)\citenamefont
  {Ariosto}, \citenamefont {Pacelli}, \citenamefont {Pastore}, \citenamefont
  {Ginelli}, \citenamefont {Gherardi},\ and\ \citenamefont
  {Rotondo}}]{ariosto2022statistical}%
  \BibitemOpen
  \bibfield  {author} {\bibinfo {author} {\bibfnamefont {S.}~\bibnamefont
  {Ariosto}}, \bibinfo {author} {\bibfnamefont {R.}~\bibnamefont {Pacelli}},
  \bibinfo {author} {\bibfnamefont {M.}~\bibnamefont {Pastore}}, \bibinfo
  {author} {\bibfnamefont {F.}~\bibnamefont {Ginelli}}, \bibinfo {author}
  {\bibfnamefont {M.}~\bibnamefont {Gherardi}},\ and\ \bibinfo {author}
  {\bibfnamefont {P.}~\bibnamefont {Rotondo}},\ }\bibfield  {title} {\bibinfo
  {title} {Statistical mechanics of deep learning beyond the infinite-width
  limit},\ }\href@noop {} {\bibfield  {journal} {\bibinfo  {journal} {arXiv
  preprint arXiv:2209.04882}\ } (\bibinfo {year} {2022})}\BibitemShut {NoStop}%
\bibitem [{\citenamefont {Seung}\ \emph {et~al.}(1992)\citenamefont {Seung},
  \citenamefont {Sompolinsky},\ and\ \citenamefont
  {Tishby}}]{seung1992statistical}%
  \BibitemOpen
  \bibfield  {author} {\bibinfo {author} {\bibfnamefont {H.~S.}\ \bibnamefont
  {Seung}}, \bibinfo {author} {\bibfnamefont {H.}~\bibnamefont {Sompolinsky}},\
  and\ \bibinfo {author} {\bibfnamefont {N.}~\bibnamefont {Tishby}},\
  }\bibfield  {title} {\bibinfo {title} {Statistical mechanics of learning from
  examples},\ }\href@noop {} {\bibfield  {journal} {\bibinfo  {journal}
  {Physical review A}\ }\textbf {\bibinfo {volume} {45}},\ \bibinfo {pages}
  {6056} (\bibinfo {year} {1992})}\BibitemShut {NoStop}%
\bibitem [{\citenamefont {Li}\ and\ \citenamefont
  {Sompolinsky}(2021)}]{li2021statistical}%
  \BibitemOpen
  \bibfield  {author} {\bibinfo {author} {\bibfnamefont {Q.}~\bibnamefont
  {Li}}\ and\ \bibinfo {author} {\bibfnamefont {H.}~\bibnamefont
  {Sompolinsky}},\ }\bibfield  {title} {\bibinfo {title} {Statistical mechanics
  of deep linear neural networks: The backpropagating kernel renormalization},\
  }\href@noop {} {\bibfield  {journal} {\bibinfo  {journal} {Physical Review
  X}\ }\textbf {\bibinfo {volume} {11}},\ \bibinfo {pages} {031059} (\bibinfo
  {year} {2021})}\BibitemShut {NoStop}%
\bibitem [{\citenamefont {Lee}\ \emph {et~al.}(2020)\citenamefont {Lee},
  \citenamefont {Schoenholz}, \citenamefont {Pennington}, \citenamefont
  {Adlam}, \citenamefont {Xiao}, \citenamefont {Novak},\ and\ \citenamefont
  {Sohl-Dickstein}}]{lee2020finite}%
  \BibitemOpen
  \bibfield  {author} {\bibinfo {author} {\bibfnamefont {J.}~\bibnamefont
  {Lee}}, \bibinfo {author} {\bibfnamefont {S.}~\bibnamefont {Schoenholz}},
  \bibinfo {author} {\bibfnamefont {J.}~\bibnamefont {Pennington}}, \bibinfo
  {author} {\bibfnamefont {B.}~\bibnamefont {Adlam}}, \bibinfo {author}
  {\bibfnamefont {L.}~\bibnamefont {Xiao}}, \bibinfo {author} {\bibfnamefont
  {R.}~\bibnamefont {Novak}},\ and\ \bibinfo {author} {\bibfnamefont
  {J.}~\bibnamefont {Sohl-Dickstein}},\ }\bibfield  {title} {\bibinfo {title}
  {Finite versus infinite neural networks: an empirical study},\ }\href@noop {}
  {\bibfield  {journal} {\bibinfo  {journal} {Advances in Neural Information
  Processing Systems}\ }\textbf {\bibinfo {volume} {33}},\ \bibinfo {pages}
  {15156} (\bibinfo {year} {2020})}\BibitemShut {NoStop}%
\bibitem [{\citenamefont {Csat{\'o}}\ \emph {et~al.}(1999)\citenamefont
  {Csat{\'o}}, \citenamefont {Fokou{\'e}}, \citenamefont {Opper}, \citenamefont
  {Schottky},\ and\ \citenamefont {Winther}}]{csato1999efficient}%
  \BibitemOpen
  \bibfield  {author} {\bibinfo {author} {\bibfnamefont {L.}~\bibnamefont
  {Csat{\'o}}}, \bibinfo {author} {\bibfnamefont {E.}~\bibnamefont
  {Fokou{\'e}}}, \bibinfo {author} {\bibfnamefont {M.}~\bibnamefont {Opper}},
  \bibinfo {author} {\bibfnamefont {B.}~\bibnamefont {Schottky}},\ and\
  \bibinfo {author} {\bibfnamefont {O.}~\bibnamefont {Winther}},\ }\bibfield
  {title} {\bibinfo {title} {Efficient approaches to {Gaussian} process
  classification},\ }\href@noop {} {\bibfield  {journal} {\bibinfo  {journal}
  {Advances in neural information processing systems}\ }\textbf {\bibinfo
  {volume} {12}} (\bibinfo {year} {1999})}\BibitemShut {NoStop}%
\bibitem [{\citenamefont {Sollich}(1998)}]{sollich1998learning}%
  \BibitemOpen
  \bibfield  {author} {\bibinfo {author} {\bibfnamefont {P.}~\bibnamefont
  {Sollich}},\ }\bibfield  {title} {\bibinfo {title} {Learning curves for
  {Gaussian} processes},\ }\href@noop {} {\bibfield  {journal} {\bibinfo
  {journal} {Advances in neural information processing systems}\ }\textbf
  {\bibinfo {volume} {11}} (\bibinfo {year} {1998})}\BibitemShut {NoStop}%
\bibitem [{\citenamefont {Williams}\ and\ \citenamefont
  {Vivarelli}(2000)}]{williams2000upper}%
  \BibitemOpen
  \bibfield  {author} {\bibinfo {author} {\bibfnamefont {C.~K.}\ \bibnamefont
  {Williams}}\ and\ \bibinfo {author} {\bibfnamefont {F.}~\bibnamefont
  {Vivarelli}},\ }\bibfield  {title} {\bibinfo {title} {Upper and lower bounds
  on the learning curve for {Gaussian} processes},\ }\href@noop {} {\bibfield
  {journal} {\bibinfo  {journal} {Machine Learning}\ }\textbf {\bibinfo
  {volume} {40}},\ \bibinfo {pages} {77} (\bibinfo {year} {2000})}\BibitemShut
  {NoStop}%
\bibitem [{\citenamefont {Seeger}(2002)}]{seeger2002pac}%
  \BibitemOpen
  \bibfield  {author} {\bibinfo {author} {\bibfnamefont {M.}~\bibnamefont
  {Seeger}},\ }\bibfield  {title} {\bibinfo {title} {{PAC-Bayesian}
  generalisation error bounds for {Gaussian} process classification},\
  }\href@noop {} {\bibfield  {journal} {\bibinfo  {journal} {Journal of machine
  learning research}\ }\textbf {\bibinfo {volume} {3}},\ \bibinfo {pages} {233}
  (\bibinfo {year} {2002})}\BibitemShut {NoStop}%
\bibitem [{\citenamefont {Arora}\ \emph {et~al.}(2019)\citenamefont {Arora},
  \citenamefont {Du}, \citenamefont {Hu}, \citenamefont {Li}, \citenamefont
  {Salakhutdinov},\ and\ \citenamefont {Wang}}]{arora2019exact}%
  \BibitemOpen
  \bibfield  {author} {\bibinfo {author} {\bibfnamefont {S.}~\bibnamefont
  {Arora}}, \bibinfo {author} {\bibfnamefont {S.~S.}\ \bibnamefont {Du}},
  \bibinfo {author} {\bibfnamefont {W.}~\bibnamefont {Hu}}, \bibinfo {author}
  {\bibfnamefont {Z.}~\bibnamefont {Li}}, \bibinfo {author} {\bibfnamefont
  {R.~R.}\ \bibnamefont {Salakhutdinov}},\ and\ \bibinfo {author}
  {\bibfnamefont {R.}~\bibnamefont {Wang}},\ }\bibfield  {title} {\bibinfo
  {title} {On exact computation with an infinitely wide neural net},\
  }\href@noop {} {\bibfield  {journal} {\bibinfo  {journal} {Advances in Neural
  Information Processing Systems}\ }\textbf {\bibinfo {volume} {32}},\ \bibinfo
  {pages} {8141} (\bibinfo {year} {2019})}\BibitemShut {NoStop}%
\bibitem [{\citenamefont {Hanin}\ and\ \citenamefont
  {Nica}(2019)}]{hanin2019finite}%
  \BibitemOpen
  \bibfield  {author} {\bibinfo {author} {\bibfnamefont {B.}~\bibnamefont
  {Hanin}}\ and\ \bibinfo {author} {\bibfnamefont {M.}~\bibnamefont {Nica}},\
  }\bibfield  {title} {\bibinfo {title} {Finite depth and width corrections to
  the neural tangent kernel},\ }in\ \href@noop {} {\emph {\bibinfo {booktitle}
  {International Conference on Learning Representations}}}\ (\bibinfo {year}
  {2019})\BibitemShut {NoStop}%
\bibitem [{\citenamefont {Karakida}\ \emph {et~al.}(2019)\citenamefont
  {Karakida}, \citenamefont {Akaho},\ and\ \citenamefont
  {Amari}}]{karakida2019universal}%
  \BibitemOpen
  \bibfield  {author} {\bibinfo {author} {\bibfnamefont {R.}~\bibnamefont
  {Karakida}}, \bibinfo {author} {\bibfnamefont {S.}~\bibnamefont {Akaho}},\
  and\ \bibinfo {author} {\bibfnamefont {S.-i.}\ \bibnamefont {Amari}},\
  }\bibfield  {title} {\bibinfo {title} {Universal statistics of {Fisher}
  information in deep neural networks: Mean field approach},\ }in\ \href@noop
  {} {\emph {\bibinfo {booktitle} {The 22nd International Conference on
  Artificial Intelligence and Statistics}}}\ (\bibinfo {year} {2019})\ pp.\
  \bibinfo {pages} {1032--1041}\BibitemShut {NoStop}%
\bibitem [{\citenamefont {Khan}\ \emph {et~al.}(2019)\citenamefont {Khan},
  \citenamefont {Immer}, \citenamefont {Abedi},\ and\ \citenamefont
  {Korzepa}}]{khan2019approximate}%
  \BibitemOpen
  \bibfield  {author} {\bibinfo {author} {\bibfnamefont {M.~E.~E.}\
  \bibnamefont {Khan}}, \bibinfo {author} {\bibfnamefont {A.}~\bibnamefont
  {Immer}}, \bibinfo {author} {\bibfnamefont {E.}~\bibnamefont {Abedi}},\ and\
  \bibinfo {author} {\bibfnamefont {M.}~\bibnamefont {Korzepa}},\ }\bibfield
  {title} {\bibinfo {title} {Approximate inference turns deep networks into
  {Gaussian} processes},\ }in\ \href@noop {} {\emph {\bibinfo {booktitle}
  {Advances in neural information processing systems}}}\ (\bibinfo {year}
  {2019})\ pp.\ \bibinfo {pages} {3094--3104}\BibitemShut {NoStop}%
\end{thebibliography}%

\appendix

\section{Expression for non-Gaussian prior distribution}\label{Edgeworth_Expansion}

After explicitly taking the derivatives in Eq.~(\ref{Edgeworth_Joint}), it is straightforward to show the expressions for the multivariate non-Gaussian distribution reads,
\begin{equation}
	q({\bf z})=\frac{1}{\sqrt{(2\pi)^N|K|}}e^{-\frac{1}{2}{\bf z}^tK^{-1}{\bf z}}\bigg[1+
	\frac{V_{\alpha\beta\gamma\delta}}{24}(\chi^{(0)}_{\alpha\beta\gamma\delta}
	-\chi^{(1)}_{\alpha\beta\gamma\delta\mu\nu}z_{\mu}z_{\nu}
	+\chi^{(2)}_{\alpha\beta\gamma\delta\mu\nu\kappa\eta}z_{\mu}z_{\nu}z_{\kappa}z_{\eta})
	\bigg]\:,
\end{equation} in which the first perturbation term, a scalar term, is the contraction between the fourth cumulant tensor $V_{\alpha\beta\gamma\delta}\propto1/N_1$ and the following tensor,
\begin{equation}
	\chi^{(0)}_{\alpha\beta\gamma\delta} = (K^{-1})_{\alpha\beta}(K^{-1})_{\gamma\delta}
	+(K^{-1})_{\alpha\gamma}(K^{-1})_{\beta\delta}
	+(K^{-1})_{\alpha\delta}(K^{-1})_{\beta\gamma}\:. 
\end{equation} The next contribution results from contraction between the fourth cumulant, the following rank-6 tensor,
\begin{equation}
\begin{split}
	\chi^{(1)}_{\alpha\beta\gamma\delta\mu\nu} = 
	&(K^{-1})_{\alpha\mu}(K^{-1})_{\nu\beta}(K^{-1})_{\gamma\delta}
	+(K^{-1})_{\alpha\beta}(K^{-1})_{\gamma\mu}(K^{-1})_{\nu\delta}\ +\\
	&(K^{-1})_{\alpha\mu}(K^{-1})_{\nu\gamma}(K^{-1})_{\beta\delta}
	+(K^{-1})_{\alpha\gamma}(K^{-1})_{\beta\mu}(K^{-1})_{\nu\delta}\ +\\
	&(K^{-1})_{\alpha\mu}(K^{-1})_{\nu\delta}(K^{-1})_{\beta\gamma}
	+(K^{-1})_{\alpha\delta}(K^{-1})_{\beta\mu}(K^{-1})_{\nu\gamma} \:,
\end{split}
\end{equation}and the rank-2 tensor $z_{\mu}z_{\nu}$. The last contribution contains the rank-8 tensor,
\begin{equation}
	\chi^{(2)}_{\alpha\beta\gamma\delta\mu\nu\kappa\eta} 
	= (K^{-1})_{\alpha\mu}(K^{-1})_{\beta\nu}(K^{-1})_{\gamma\kappa}(K^{-1})_{\delta\eta}\:,
\end{equation} the rank-4 tensor $z_{\mu}z_{\nu}z_{\kappa}z_{\eta}$ and the fourth cumulant. It becomes extensively tedious, if possible, proving the marginal and conditional properties with this representation.

\section{Proofs of conditional mean Eq.~(\ref{pred_mean}) and second moment Eq.~(\ref{pred_2nd_moment})}\label{cond-mean}
By plugging the corresponding expression for non-Gaussian joint $q(z_1,\tilde{\bf z})$, using the conditional property of distribution, and temporarily neglecting the coefficients and indices related to the fourth cumulant $V$, we can compute the conditional first moment in the followings,
\begin{equation}
\begin{split}
	\mathbb E[z_1|\tilde{\bf z}] &\propto \int dz_1 z_1q(z_1,\tilde{\bf z}) \\
	&=\int dz_1z_1(1+V\tilde\partial^4 + V\tilde\partial^3\partial_{z_1} +
	V\tilde\partial^2\partial^2_{z_1} + V\tilde\partial\partial^3_{z_1} + V\partial^4_{z_1})
	[\mathcal N(z_1|\tilde{\bf z})\mathcal N(\tilde{\bf z})]\\
	&=(1+V\tilde\partial^4)[\mu(\tilde{\bf z})\mathcal N(\tilde{\bf z})]
	-V\tilde\partial^3\mathcal N(\tilde{\bf z})\:.
\end{split}
\end{equation} In the first equality, the derivative terms of equal to or higher order than $\partial^2_{z_1}$ do not contribute as they produce zero when acting on $z_1$. In the second equality, one can further reduce the first term into %$\tilde\partial^4(\mu\mathcal N)=
$\mu(1+V\tilde\partial^4)\mathcal N+4(\tilde\partial\mu)\tilde\partial^3\mathcal N$ since the $\mu$ term is linear in $\tilde{\bf z}$. Note that the former term is actually $\mu q(\tilde{\bf z})$. In addition, the minus sign preceding the $\tilde\partial^3$ arises from the adjoint property of moving $\partial_{z_1}$ to act on $z_1$. By restoring the coefficients with some combinatorics and that $\partial_i\mu({\bf z})=[K^{-1}{\bf k}]_i$, it proves Eq.~(\ref{pred_mean}).%In addition, the 

The conditional second moment can be computed in a similar manner,
\begin{equation}
\begin{split}
	\mathbb E[z_1^2|\tilde{\bf z}] &\propto \int dz_1 z_1^2q(z_1,\tilde{\bf z}) \\
	&=\int dz_1z_1^2(1+V\tilde\partial^4 + V\tilde\partial^3\partial_{z_1} +
	V\tilde\partial^2\partial^2_{z_1} + V\tilde\partial\partial^3_{z_1} + V\partial^4_{z_1})
	[\mathcal N(z_1|\tilde{\bf z})\mathcal N(\tilde{\bf z})]\\
	&=(1+V\tilde\partial^4)\bigg\{
	[\mu^2(\tilde{\bf z})+\sigma^2]\mathcal N(\tilde{\bf z})
	\bigg\}
	-2V\tilde\partial^3[\mu(\tilde{\bf z})\mathcal N(\tilde{\bf z})]
	+2V\tilde\partial^2[\mathcal N(\tilde{\bf z})]\:.
\end{split}
\end{equation} The first term in the second equality shall contribute a term $\propto q(\tilde{\bf z})$, the the details are similar with above.

\section{Proofs of evidence in Eq.~(\ref{evidence}) and posterior in Eq.~(\ref{bayes_posterior})}\label{evidence_proof}
We start with the definition in Eq.~(\ref{evidence}) and focus on the correction term. The computation involves the marginalization over all $z$'s as well as a complication arising from the derivatives with respect to some 4 $z$'s. 
%Nevertheless, the expression can be simplified by moving all the derivatives out of the integral. 
Using the adjoint property of derivative, i.e. $\int f(x)[\partial g(x)/\partial x]dx = \int[-\partial f(x)/\partial x]g(x)dx$, provided that the product $fg$ vanishes at infinity, we can shuffle the operator $\partial_z$ acting on the Gaussian prior on the right most to act on the Gaussian likelihood on the left. Then use the property, $(\partial_t+\partial_s)\exp[-(s-t)^2]=0$, we can replace $\partial_z$ with $-\partial_y$, and pull these derivatives out of the integral. The followings summarize these actions. %are summarized in the followings, %We can proceed by firstly decomposing the observed values ${\bf y}={\bf y}'\cup\{y_i,y_j,y_l,y_m\}$ and similarly the function values ${\bf z}={\bf z}'\cup\{z_i,z_j,z_l,z_m\}$. So ${\bf y}':={\bf y}\backslash\{y_i,y_j,y_l,y_m\}$ and likewise for ${\bf z}'$. Such decomposition 
\begin{equation}
\begin{split}
	&\int d{\bf z}\ \mathcal N({\bf y}|{\bf z},\Lambda)\big[ 
	v_{ijml}\partial_{z_i}\partial_{z_j}\partial_{z_m}\partial_{z_l}\mathcal N({\bf z}|0,K)\big]\\
	%=v_{ijml}
	%\int d{\bf z}'%'dz_idz_jdz_mdz_l
	%\mathcal N({\bf y}'|{\bf z}',\Lambda')\mathcal N({\bf z}'|0,K')\\
	%&\times\int dz_{i,j,m,l}\mathcal N(y_{i,j,m,l}|z_{i,j,m,l},\Lambda_{4\times4})
	%\partial^{(4)}_{z_{i,j,m,l}}%\partial_{z_j}\partial_{z_l}\partial_{z_m}
	%\mathcal N(z_{i,j,m,l}|{\bf z}')\\
	&=v_{ijml}\int d{\bf z}%'dz_idz_jdz_mdz_l
	\big[
	\partial_{z_i}\partial_{z_j}\partial_{z_m}\partial_{z_l}
	\mathcal N({\bf y}|{\bf z},\Lambda)
	\big]\mathcal N({\bf z}|0,K)\\
	%\int dz_{i,j,m,l}\big[ 
	%\partial^{(4)}_{y_{i,j,m,l}}\mathcal N(y_{i,j,m,l}|z_{i,j,m,l},\Lambda_{4\times4})
	%\big]\mathcal N(z_{i,j,m,l}|{\bf z}')\\
	&=v_{ijml}\partial_{y_i}\partial_{y_j}\partial_{y_m}\partial_{y_l}
	\int d{\bf z}\mathcal N({\bf y}|{\bf z},\Lambda)\mathcal N({\bf z}|0,K)\\
	&=v_{ijml}\partial_{y_i}\partial_{y_j}\partial_{y_m}\partial_{y_l}
	\mathcal N({\bf y}|0,K+\Lambda)\:.
\end{split}
\end{equation} %where in reaching the second equality we have exploited the fact that $(\partial_t+\partial_s)\exp[-(s-t)^2]=0$ to replace $\partial_z$ with $-\partial_y$ so they can be factored out of the integral.
As for the proof for Eq.~(\ref{bayes_posterior}), we can proceed with computation of the numerator there, and note that the derivative $\partial_{z_*}$ in $q(z_*,{\bf z})$ can be moved out of the integral directly. 
%under different situation which does not appear previously. In addition, $\partial_{z_*}$ can be moved out of the integration directly. 
To accommodate this extra variable, we use the hat on top of the indices to stress that $*$ is included. The followings contain the meaning of our notation as well as some details. 
\begin{equation}
\begin{split}
v_{\hat i\hat j\hat m\hat l}&\int d{\bf z}\ \mathcal N({\bf y}|{\bf z},\Lambda)\big[ 
	\partial_{z_{\hat i}}\partial_{z_{\hat j}}\partial_{z_{\hat m}}\partial_{z_{\hat l}}\mathcal N(z_*,{\bf z}|0,\hat K)
	\big]\\
	&=\bigg[
	v_{ijml}\partial_{y_i}\partial_{y_j}\partial_{y_m}\partial_{y_l}+
	v_{ijm*}\partial_{y_i}\partial_{y_j}\partial_{y_m}\partial_{z_*}+
	v_{ij**}\partial_{y_i}\partial_{y_j}\partial^2_{z_*}+
	v_{i***}\partial_{y_i}\partial^3_{z_*}+
	v_{****}\partial^4_{z_*}
	\bigg]\times\\
	%&\int d{\bf z}\ \mathcal N(z_*|{\bf k}^t_*K^{-1}{\bf z},k_{**}-{\bf k}^t_*K^{-1}{\bf k}_*)
	%\mathcal N({\bf z}|K(K+\Lambda)^{-1}{\bf y}, \Lambda K(K+\Lambda)^{-1})
	%\mathcal N({\bf y}|0,K+\Lambda)\\
	&\int d{\bf z} \mathcal N({\bf y}|{\bf z},\Lambda) \mathcal N(z_*,{\bf z}|0,\hat K)\\
	&=v_{\hat i\hat j\hat m\hat l}\partial_{\hat i\hat j\hat m\hat l}%\bigg[
	\mathcal N\big[\bigl(\begin{smallmatrix}z_* \\{\bf y} \end{smallmatrix}\bigr)
	\big|0,\bigl(
	\begin{smallmatrix}1 &{\bf k}_*^t\\{\bf k}_*&K+\Lambda \end{smallmatrix}\bigr)
	%z_*|{\bf k}^t_*(K+\Lambda)^{-1}{\bf y},k_{**}-{\bf k}^t_*(K+\Lambda)^{-1}{\bf k}_*
	\big]\:,
	%\mathcal N({\bf y}|0,K+\Lambda)
	%\bigg]
\end{split}
\end{equation} where the readers can refer to the text around Eq.~(2.21) in~\cite{rasmussen2006gaussian} for similar integration leading to the second equality.
%The covariance of the two-layer cosine network can be computed as,

%\section{Generalized fourth order Hermite polynomials for bivariate distribution}
%Denoting the covariance matrix $\Sigma=\begin{bmatrix}1 & c\\c&1\end{bmatrix}$ for the bivariate normal distribution $\mathcal N(y_1,y_2)$, we can define the corresponding fourth order Hermite polynomial as,
%\begin{equation}
%\begin{split}
%	He_{4,0}(y_1,y_2)&:=[\mathcal N(y_1,y_2|0,\Sigma)]^{-1}\partial^4_{y_1}\mathcal N(y_1,y_2|0,\Sigma)\\
%	&=\frac{3}{(1-c^2)^2}-\frac{6(y_1-cy_2)^2}{(1-c^2)^3}%-\frac{6(y_1-cy_2)^2}{(1-c^2)^3}
%	+\frac{(y_1-cy_2)^4}{(1-c^2)^4}\:.
%\end{split}
%\end{equation} In the case $c=0$, $He_{4,0}(y_1,y_2)$ coincides with the ordinary Hermite polynomial $He_4(y_1)$. Similarly, one can use the fact $\partial/\partial y_1=\partial/\partial(-cy_2)$ when applying to the bivariate normal distribution, and 
%\begin{equation}
%\begin{split}
%	He_{3,1}(y_1,y_2)&=(-c)He_{4,0}(y_1,y_2)\\
	%\frac{-3c}{(1-c^2)^2}+\frac{6c(y_1-cy_2)^2}{(1-c^2)^3}%-\frac{6(y_1-cy_2)^2}{(1-c^2)^3}
	%-\frac{3c(y_1-cy_2)^4}{(1-c^2)^4}\\
%	He_{2,2}(y_1,y_2)&=c^2He_{4,0}(y_1,y_2)
	%\frac{3c^2}{(1-c^2)^2}-\frac{6c^2(y_1-cy_2)^2}{(1-c^2)^3}%-\frac{6(y_1-cy_2)^2}{(1-c^2)^3}
	%+\frac{c^2(y_1-cy_2)^4}{(1-c^2)^4}
%\end{split}
%\end{equation}

\section{Proof of Eq.~(\ref{dgp_2nd_moment})}\label{2nd_moment_correction}
To prove the correction term to the second moment, the previous tricks dealing with Gaussians are still useful. In below, we first shuffle the four derivatives acting on the bivariate Gaussian on the left most to act on the Gaussian factor. One can further use $\partial_{z_{1,2}}\exp[-(z_1-z_2)^2/2]=\pm \partial_{z_-}\exp(-z_-^2/2)$ such that one can rewrite $\partial_{ijlm}=(-1)^{n_1}\partial^4_{z_-}$ with $n_1$ denoting the number of involved $z_1$'s. The rest of computation is straightforward.
%\cite{lu2018standing}
\begin{equation}
\begin{split}
	&\int dz_1dz_2\ e^{-\frac{(z_1-z_2)^2}{2H}}(v_{ijml}\partial_i\partial_j\partial_m\partial_l)
	\mathcal N(z_1,z_2|0,K_2)\\
	&= \int dz_1dz_2\big [
	(v_{ijml}\partial_i\partial_j\partial_m\partial_l)e^{-\frac{(z_1-z_2)^2}{2H}}
	\big]\mathcal N(z_1,z_2|0,K_2)\\
	&= \frac{V_{1111}+V_{2222}-4V_{1222}-4V_{21111}+6V_{1122}}{24}\times\\
	&\int \frac{dz_+dz_-}{2}[(\partial_{z_-})^4e^{-\frac{z_-^2}{2H}}]
	\mathcal N(z_+|0,\sigma_+^2)\mathcal N(z_-|0,\sigma_-^2)\\
	&=V'\int dz 
	(\frac{3}{H^2}-\frac{6z^2}{H^3}+\frac{z^4}{H^4})
	e^{-\frac{z^2}{2H}}\mathcal N(z|0,\sigma_-^2)\\
	&=\frac{V'}{\sqrt{1+\frac{\sigma_-^2}{H}}}
	\big[\frac{3}{H^2}
	-\frac{6\sigma_-^2}{H^3(1+\sigma_-^2/H)}
	+\frac{3\sigma_-^4}{H^4(1+\sigma_-^2/H)^2}\big]\:.
\end{split}
\end{equation} We have used the notation for the variances $\sigma_{\pm}^2=2[1\pm k({\bf x}_1,{\bf x}_2)]$.

\end{document}